\documentclass[acmlarge]{acmart}
\AtBeginDocument{%
  }

\usepackage{listings} 
\usepackage{array} 
\usepackage{xcolor}
\usepackage{adjustbox}

\definecolor{codegray}{rgb}{0.95,0.95,0.95}
\definecolor{commentgray}{rgb}{0.4,0.4,0.4}
\definecolor{keywordblue}{rgb}{0.2,0.2,0.7}
\definecolor{stringred}{rgb}{0.8,0.1,0.1}
\definecolor{linenumbergray}{rgb}{0.5,0.5,0.5}
\definecolor{framegray}{rgb}{0.75,0.75,0.75}

\lstdefinelanguage{Solidity}{
  morekeywords={
    contract,function,modifier,event,enum,struct,if,else,while,for,import,return,
    mapping,address,bool,string,public,private,internal,external,view,pure,storage,memory,new,
    require,assert,revert,emit,calldata,override,virtual,constructor
  },
  sensitive=true,
  morecomment=[l]{//},
  morecomment=[s]{/*}{*/},
  morestring=[b]",
}

\begin{document}

\title{A Comprehensive Study of Exploitable Patterns in Smart Contracts: From Vulnerability to Defense}


\author{Yuchen Ding}
\authornote{Yuchen Ding and Hongli Peng contributed equally to this work.}
\affiliation{%
  \institution{Hainan University}
  \city{Haikou}
  \state{Hainan}
  \country{China}
}

\author{Hongli Peng}
\authornotemark[1] 
\affiliation{%
  \institution{Hainan University}
  \city{Haikou}
  \state{Hainan}
  \country{China}
}

\author{Xiaoqi Li}
\email{csxqli@hainanu.edu.cn}
\affiliation{%
  \institution{Hainan University}
  \city{Haikou}
  \state{Hainan}
  \country{China}
}







\renewcommand{\shortauthors}{Yuchen Ding et al.}

\begin{abstract}

With the rapid advancement of blockchain technology, smart contracts have enabled the implementation of increasingly complex functionalities. However, ensuring the security of smart contracts remains a persistent challenge across the stages of development, compilation, and execution. Vulnerabilities within smart contracts not only undermine the security of individual applications but also pose significant risks to the broader blockchain ecosystem, as demonstrated by the growing frequency of attacks since 2016, resulting in substantial financial losses. This paper provides a comprehensive analysis of key security risks in Ethereum smart contracts, specifically those written in Solidity and executed on the Ethereum Virtual Machine (EVM). We focus on two prevalent and critical vulnerability types—reentrancy and integer overflow—by examining their underlying mechanisms, replicating attack scenarios, and assessing effective countermeasures.

\end{abstract}




\maketitle

\section{Introduction}
The concept of smart contracts can be traced back to the 1990s. However, as blockchain technology has been increasingly applied across various domains, its inherent security shortcomings have become more apparent. Several notable attacks have demonstrated the severity of these vulnerabilities, including the 2016 DAO attack (reentrancy attack), the 2017 Parity attack (involving unprotected self-destruct and delegatecall misuse), and the 2018 BEC integer overflow exploit. Each of these incidents resulted in significant financial losses, highlighting the urgent need for improved security measures.  

Despite the continuous evolution of blockchain technology, security protection for smart contracts remains in its early stages. According to relevant studies, smart contract security research incorporating traditional cryptographic techniques only began to emerge in 2016 following the DAO attack. Subsequently, in 2017, research on smart contract security started gaining momentum and has since expanded rapidly \cite{hu2021research}.

In 2021, the domestic security team SharkTeam, in collaboration with OWASP, conducted a live-sharing session focused on smart contract security. This initiative involved automated security scans of mainstream blockchain projects, followed by comprehensive vulnerability analysis. In recent years, security research on blockchain smart contracts has gained increasing attention and widespread adoption \cite{jeon2024design,he2023detection,li2024cobra,liu2025sok,bu2025smartbugbert,li2024stateguard,li2024scla}. The growing interest from security professionals and organizations underscores the fact that smart contract security has become a critical concern for the continued development of blockchain technology \cite{kuo2023,subramanian2022,qi2023,zou2025malicious,mao2024scla,li2021hybrid,li2017discovering,li2025scalm}.

Blockchain technology has gradually permeated various industries, emerging as a transformative force in redefining industry standards. Among its most promising applications is the development of smart contracts. In essence, a smart contract is a digitally defined agreement implemented through computer code that executes contractual terms automatically. Given the critical role and strong enforceability of smart contracts in transactions, it becomes evident that constructing flawless code is a formidable challenge \cite{kong2024characterizing,niu2024unveiling,liu2025sok}. Drafting a smart contract requires aligning transaction requirements with computational logic while striving for a robust and comprehensive design. However, achieving entirely vulnerability-free code is nearly impossible, as unforeseen security flaws may arise \cite{li2021clue,wang2024smart}. This underscores the necessity of dedicated security research to enhance the security of blockchain-based smart contracts \cite{li2024cobra}.  

Smart contract security primarily encompasses contract security and privacy security. Privacy security pertains to attacks targeting sensitive transaction data and the identities of involved parties \cite{hu2021research}. This paper focuses specifically on vulnerabilities associated with smart contract security, analyzing their implications and mitigation strategies.

Blockchain-based smart contracts exhibit key characteristics such as decentralization, automated execution, and real-time updates \cite{bi2018blockchain}. However, their execution is often prone to unforeseen vulnerabilities.  

(1) Based on the Solidity programming language and EVM mechanisms, smart contract vulnerabilities can be categorized into three main types:  

   (i). Solidity-Level Vulnerabilities: Reentrancy attacks, gas exhaustion \cite{liu2024gastrace}, improper state manipulation, etc .
   
   (ii). EVM Bytecode-Level Vulnerabilities: Short address attacks, stack limit issues, integer overflows, and Ether loss.
   
  (iii). Blockchain-Level Vulnerabilities: Time dependency, unpredictable states, and related issues.  

(2) According to Ethereum’s four-layer smart contract architecture, security vulnerabilities can be classified as follows:  

 (i). Application Layer Vulnerabilities: Reentrancy attacks, integer overflows, unprotected self-destruct, short address attacks, and time dependency issues \cite{bi2018blockchain}.  
   
  (ii). Data Layer Vulnerabilities: Issues related to state trie manipulation, such as empty account attacks.  
   Consensus Layer Vulnerabilities: Denial-of-service (DoS) through block stuffing, probabilistic finality issues, and validator dilemmas.  
   
  (iii). Network Layer Vulnerabilities: Peer selection attacks, exposed RPC APIs, and unlimited node creation exploits.

This paper makes the following key contributions:  

(1) We conduct an in-depth analysis of smart contract vulnerabilities based on Ethereum’s Solidity language and EVM. Specifically, we examine three major types of vulnerabilities—reentrancy attacks, 
and integer overflow attacks—by analyzing their underlying mechanisms, reproducing them through code implementation, and proposing effective security countermeasures.  

(2) Additionally, we further investigate security risks arising from the misuse of the call(), providing a comprehensive analysis and mitigation strategies.

\section{Background}

\subsection{Blockchain}

Blockchain was initially introduced as the underlying technology of Bitcoin \cite{2008bitcoin}, designed as a chain-structured distributed ledger to realize the vision of decentralization. With advancements in smart contracts and cryptographic techniques, blockchain technology has evolved into a new phase of development.

In a broader sense, blockchain can be defined as an innovative distributed infrastructure characterized by a chained structure, leveraging consensus algorithms, cryptographic methods, and automated smart contracts for secure data storage, updating, and transmission \cite{wang2016blockchain}.

\subsection{Smart Contract}
\subsubsection{Evolution of Smart Contracts}
The concept of smart contracts was first introduced by interdisciplinary scholar Nick Szabo, who defined a smart contract as “a set of digitally defined commitments, including protocols on which the involved parties can execute these commitments.” BaiduEncyclopedia. Notably, this concept predates the emergence of blockchain technology. However, the widespread adoption of smart contracts was initially constrained by the lack of a trusted execution environment and the necessary programming technologies and systems to support contract implementation.

In 2008, the advent of Bitcoin—the first cryptocurrency—marked the realization of decentralized currency. As blockchain gained recognition as Bitcoin’s underlying technology, research on its potential applications expanded. To enable more complex functionalities, the concept of smart contracts was revisited. However, due to the limitations imposed by blockchain’s rigid structure and the challenges associated with hard forks, smart contracts struggled to integrate seamlessly into existing blockchain networks. It was not until July 30, 2015, with the official launch of Ethereum’s first-generation platform, that these challenges were addressed, ushering in the Smart Contract 2.0 era \cite{bu2025enhancing}.

\subsubsection{Characteristics of Smart Contracts}
Smart contracts exhibit several key characteristics, including trustlessness, automation, immutability, and traceability. Their execution logic and predefined rules are set in advance, ensuring that once deployed, they cannot be altered by any single party. Upon activation, smart contracts self-execute and self-verify without requiring external intervention. Furthermore, by integrating blockchain’s cryptographic digital signatures and timestamp mechanisms, smart contracts enhance security and provide robust traceability.

\subsubsection{Background Knowledge on Smart Contracts}

we introduce some background knowledge on smart contracts, such as Ethereum, Solidity language, and EVM.

(1)\textbf{Ethereum}

\begin{figure}
    \centering
    \includegraphics[width=0.9\linewidth]{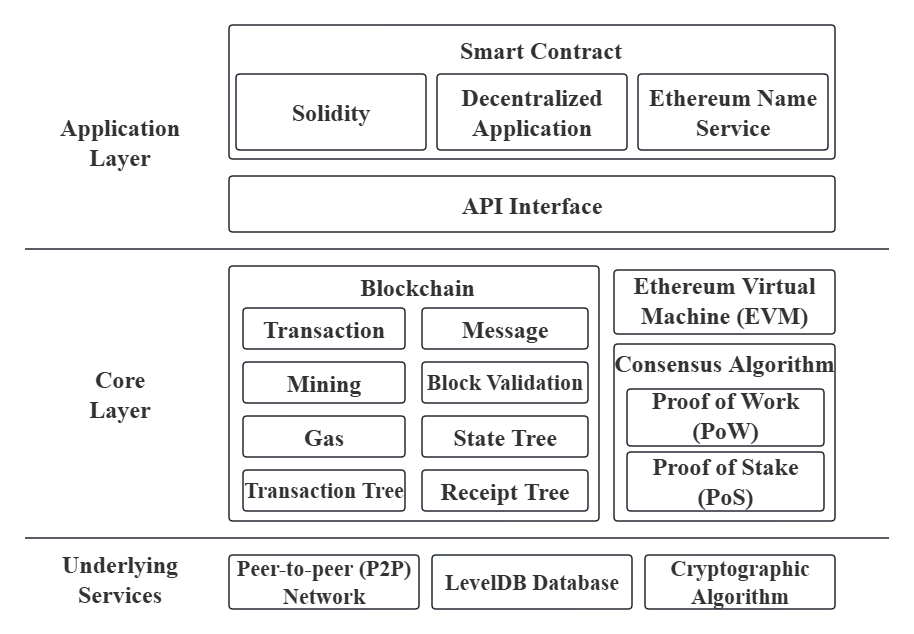}
    \caption{The Structure of Ethereum}
    \label{fig:structure of Ethereum}
    \vspace{-1em}
\end{figure}

While Bitcoin successfully demonstrated the feasibility and security of blockchain as a decentralized currency system, it suffers from limited extensibility. Its simplistic scripting language and lack of general-purpose programmability make it insufficient for supporting more advanced applications. Ethereum was specifically designed to address these limitations. As a next-generation blockchain platform, its primary objective is to provide a more flexible and programmable infrastructure, enabling the development and execution of complex decentralized applications (DApps) through smart contracts. The structure of Ethereum is shown in Figure \ref{fig:structure of Ethereum}.

\begin{figure}
    \centering
    \includegraphics[width=0.9\linewidth]{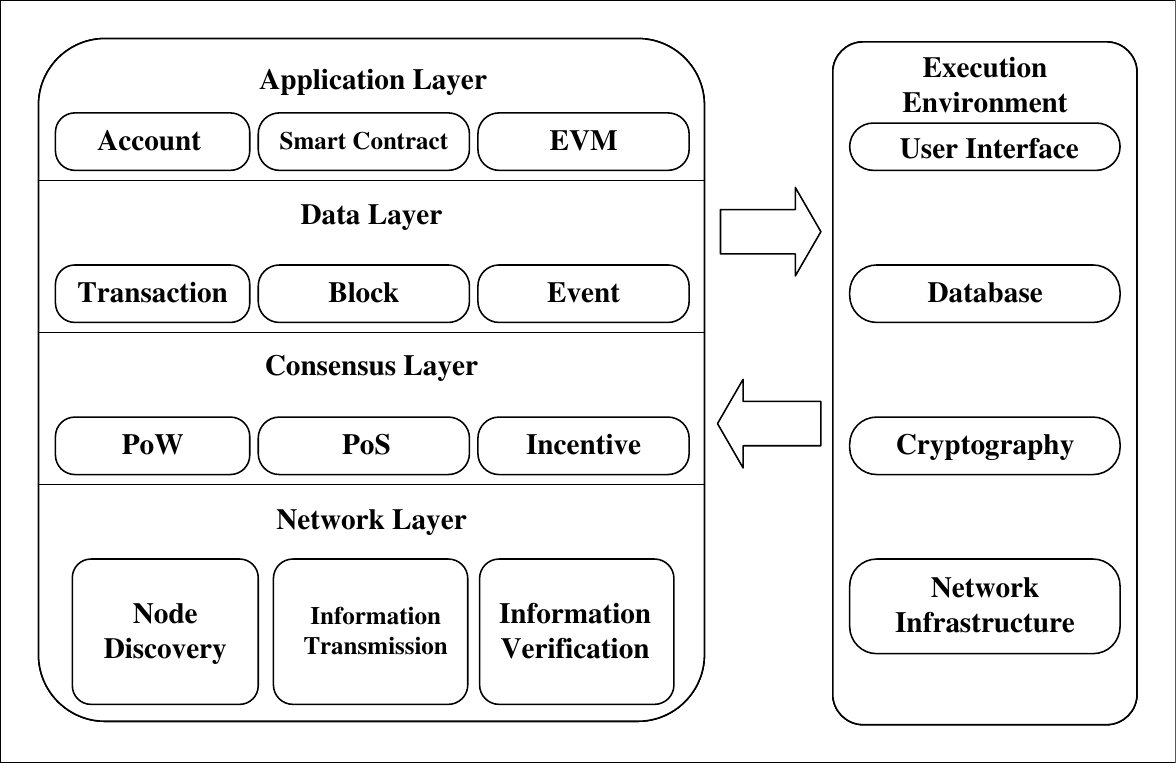}
    \caption{The Ethereum Smart Contract Architecture \cite{garba2021defi}}
    \label{fig:structure of Ethereum Smart Contract}
    \vspace{-1em}
\end{figure}

Ethereum is an open-source blockchain platform that enables anyone to create and run decentralized applications. It introduces the EVM, a Turing-complete scripting environment comparable to an assembly language, which allows high-level programming languages to be compiled into EVM bytecode for application development and execution. The emergence of Ethereum provided a programmable runtime environment and an operational platform for smart contracts, effectively integrating blockchain technology with smart contract functionality. This significantly lowered the barrier to decentralized application (DApp) development and broadened blockchain's applicability. The Ethereum smart contract architecture is shown in Figure \ref{fig:structure of Ethereum Smart Contract}.

The Ethereum smart contract architecture is structured into four main layers: the Application Layer, Data Layer, Consensus Layer, and Network Layer.
The Application Layer is responsible for the deployment and execution of smart contracts.
The Data Layer manages the underlying blockchain data structures.
The Consensus Layer ensures state consistency across the blockchain through consensus algorithms and cryptographic mechanisms.
The Network Layer facilitates peer-to-peer (P2P) communication, node synchronization, and overall network maintenance.
Ethereum-based environments typically interact with users through a web-based user interface, integrated with this four-layered architecture to support blockchain operation and smart contract execution.

(2) \textbf{Solidity Language}

Smart contracts on Ethereum are built upon EVM bytecode. However, developers typically do not write bytecode directly. Instead, they use high-level, contract-oriented languages such as Solidity to develop smart contracts.

Solidity is a high-level programming language specifically designed for writing smart contracts, with syntax and semantics similar to those of languages like C, Python, and JavaScript. Contracts written in Solidity are compiled into EVM bytecode, which is then executed within the Ethereum Virtual Machine.
Once deployed, smart contracts operate according to predefined business logic and are capable of handling various application-specific tasks autonomously on the Ethereum blockchain.

(3)\textbf{EVM}

\begin{figure}
    \centering
    \includegraphics[width=0.7\linewidth]{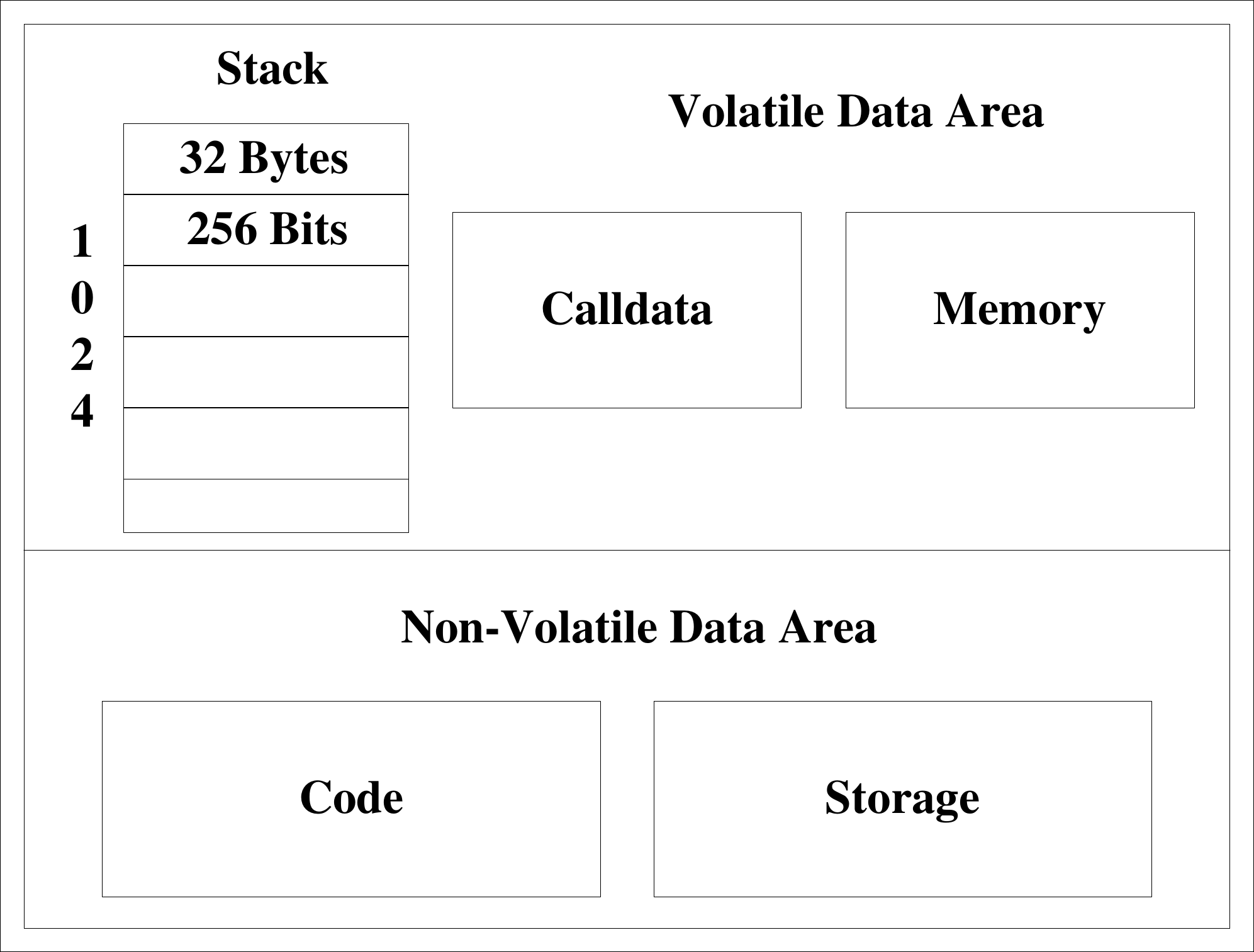}
    \caption{The Structure of EVM}
    \label{fig:structure of EVM}
    \vspace{-1em}
\end{figure}

The EVM architecture is shown in the Figure \ref{fig:structure of EVM}. The EVM consists of two main components: a volatile data area and a non-volatile data area. When executing, the EVM bytecode compiled from Solidity source code reads from the code section located in the non-volatile area of the EVM.  
The EVM is stack-based and includes a stack structure with 1,024 stack items, each consisting of 32 bytes (256 bits). This stack limitation is one of the factors that constrain the data-handling capabilities of the Solidity language.

\section{Vulnerability Analysis and Security Pattern Design}
\subsection{Reentrancy Vulnerability}

\subsubsection{Vulnerability Principle}
Ethereum smart contracts are capable of invoking external contracts and handling Ether transactions. Such external calls can be exploited by malicious actors to hijack the control flow, resulting in the execution of additional code, including recursive callbacks to the original contract, which is known as a reentrancy attack. The vulnerability was first observed in the DAO incident, which ultimately led to a hard fork of the Ethereum blockchain. In 2023, the DeFi protocol dForce suffered a reentrancy attack that exploited a vulnerability in its smart contract execution logic, enabling the attacker to repeatedly withdraw funds before the state was properly updated, ultimately resulting in a financial loss of approximately \$3.6 million. This incident exemplifies how insufficient reentrancy protections in smart contracts can be leveraged for large-scale exploits, reinforcing the urgent need for rigorous formal verification and secure coding practices in decentralized finance ecosystems.\cite{DForceattack}

\begin{figure}
    \centering
    \includegraphics[width=0.85\linewidth]{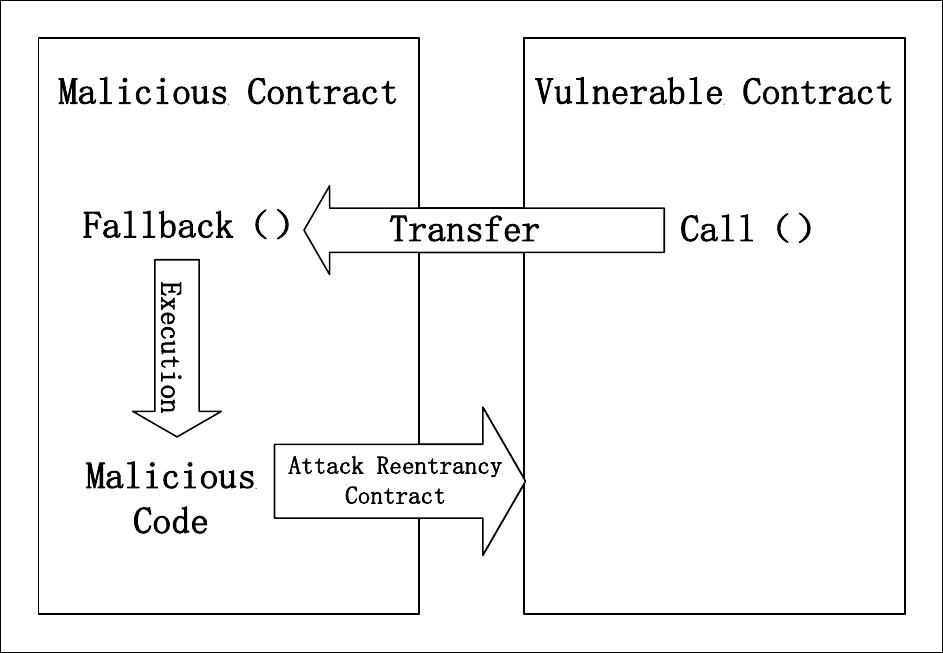}
    \caption{Reentrancy Attack Process}
    \label{fig:/reentrancy attack}
    \vspace{-1em}
\end{figure}

The reentrancy attack process is shown in Figure \ref{fig:/reentrancy attack}. To accept Ether, contracts must implement a fallback function. Without it, an exception is thrown during the Ether transfer, causing the transaction to revert. If the fallback function is embedded with malicious logic, it can be used to reenter the vulnerable contract and repeatedly execute specific functions, enabling arbitrary reentrancy attacks.

\subsubsection{Code Analysis}

We analyze the following two code examples:

\textbf{(1) Functional Implementation}

To illustrate the reentrancy vulnerability, we implement a simplified proof-of-concept consisting of two contracts:

The Bank contract is designed to record deposits from senders and facilitate fund transfers. During a transfer, it enforces strict conditions: the transfer amount must not exceed the sender’s deposited balance within the contract, nor surpass the contract’s current balance. Upon passing these checks, the contract uses the call function to transfer Ether, and correspondingly reduces the sender’s recorded balance in the contract.

The Attacker contract serves as an external interacting contract in the experiment, capable of both depositing to and withdrawing from the Bank contract.

Before initiating the attack, the file containing the Bank contract must be imported, and the target contract address should be initialized within the attacker contract as target.

To deposit funds, the statistis function in the Bank contract is invoked, while the attack function is used to trigger the withdrawal process.

Since receiving Ether is required during the attack, a fallback function must be defined. In this process, the fallback function is crafted with malicious logic to perform a reentrancy attack on the withdraw function of the Bank contract.

\textbf{(2) Attack Execution Analysis}

The Bank contract functions as a centralized bank where multiple accounts, including the Attacker, have deposited funds. Suppose the Bank contract holds a total balance of 12 ether and 2 wei, with the Attacker having deposited 2 ether.

To execute a withdrawal of 1 ether, the following steps occur:

(i). The Attacker calls the withdraw function and specifies the withdrawal amount of 1 ether, prompting the Bank contract to initiate the transfer.

(ii). The contract verifies the transfer conditions using a require statement. It then performs the transfer using the call method. Since call may invoke an unknown function, it defaults to executing the fallback function. At this point, the Bank transfers 1 ether, reducing its balance to 11 ether and 2 wei.

(iii). The fallback function evaluates the condition \texttt{if(9 ether 2 wei > 1 ether)} and recursively calls the withdraw function.

(iv). Because the contract state has not yet been updated to reflect the previous transfer, the recorded deposit remains unchanged. The contract balance is still greater than the withdrawal amount, allowing the require condition to pass again and triggering another call to transfer funds.

(v). This process of call $\rightarrow$ fallback $\rightarrow$ call is repeated, allowing multiple reentrant invocations. The loop continues until the Bank contract's balance is reduced to 2 wei. At this point, the condition in the fallback function fails, returning control to the Bank contract to finalize the execution by updating the state and ending the attack.

(vi). Result of the attack:

  The Bank contract’s balance decreases from 12 ether 2 wei to 2 wei.  

  The Attacker contract’s balance increases from 0 to 10 ether.  

  The Attacker’s recorded deposit value overflows and becomes an abnormally large number.

This demonstrates how repeated reentrant calls enable the attacker to drain all available Ether from the Bank contract, depending on the contract’s transfer restrictions and the construction of the malicious fallback logic.

\textbf{(3) Analysis Based on Function Usage}

(i). \texttt{msg.sender}, \texttt{msg.value}, \texttt{this.balance}, and \texttt{balances[address]}  
\texttt{msg.sender} refers to the actual caller of the function and is a global variable of type address.  
\texttt{msg.value} indicates the amount of Ether sent by the caller, represented as a global variable of type uint256.  
\texttt{this.balance} denotes the current contract's balance, where \texttt{this} refers to the executing contract context.  
\texttt{balances[address]} is defined via a mapping \texttt{mapping(address => uint256)} and returns the Ether balance associated with a given address in the contract.

(ii). payable modifier  
The payable modifier, when applied to a function, allows it to receive Ether. When applied to an address, it enables that address to accept Ether transfers.

(iii). fallback function  
A contract can have only one unnamed fallback function without parameters. It is executed when no matching function is found during a contract call. If marked as payable, the fallback function can also receive Ether.

(iv). Differences among call, transfer, and send  
call is a low-level function that returns true or false but not the execution result. When no matching function is found, call will default to executing the contract's fallback function. Ether transfers can be made using \texttt{address.call\{value: amount\}()}, where address is the recipient and amount is the Ether to send.  
\texttt{address.transfer(amount)} and \texttt{address.send(amount)} are also used for Ether transfers.  
The difference lies in gas forwarding: call forwards all remaining gas to the callee, enabling the fallback function to perform complex operations. transfer only forwards 2300 gas, limiting the fallback function to simple operations. Any Ether-related action within the fallback under transfer may exceed the gas limit and result in a failure.  
Additionally, send and call do not revert the transaction upon failure and allow the contract to continue executing subsequent operations. In contrast, transfer reverts the transaction if the transfer fails, halting execution.

Therefore, transfer is generally considered safer and more recommended for Ether transfers than send or call.

\subsubsection{Security Pattern Design}

The security mechanism design includes the following four approaches:

    \textbf{(1) Adopting the Checks-Effects-Interactions Pattern} 
    
\begin{figure}
        \centering
        \includegraphics[width=1\linewidth]{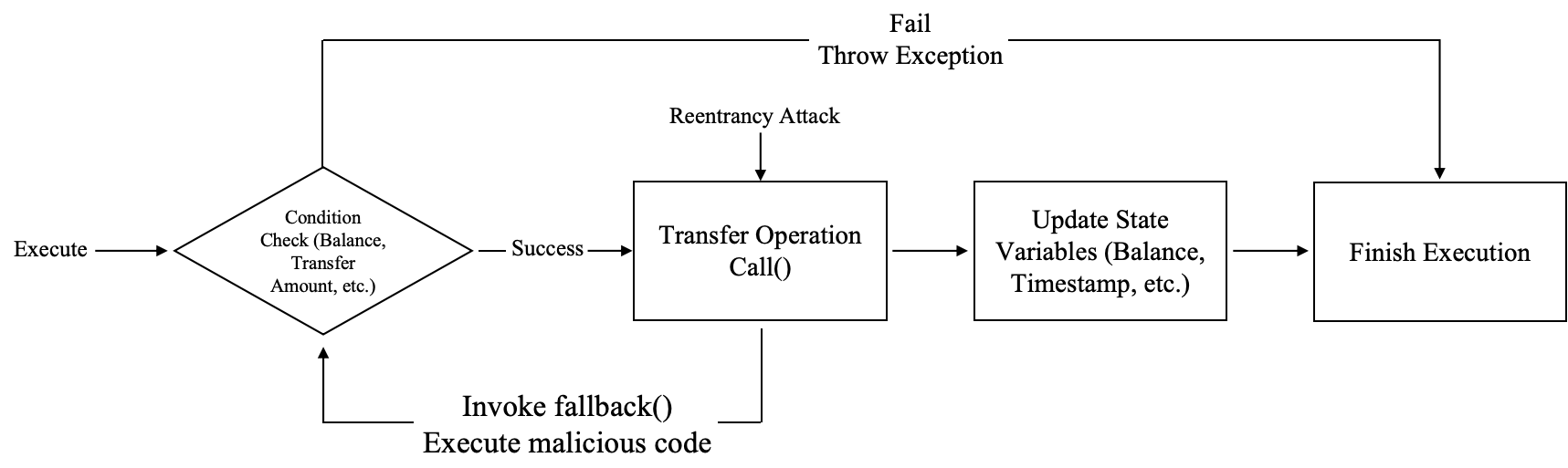}
        \caption{Vulnerability Pattern Workflow}
        \label{fig:3.2}
    \end{figure}

\begin{figure}
        \centering
        \includegraphics[width=1\linewidth]{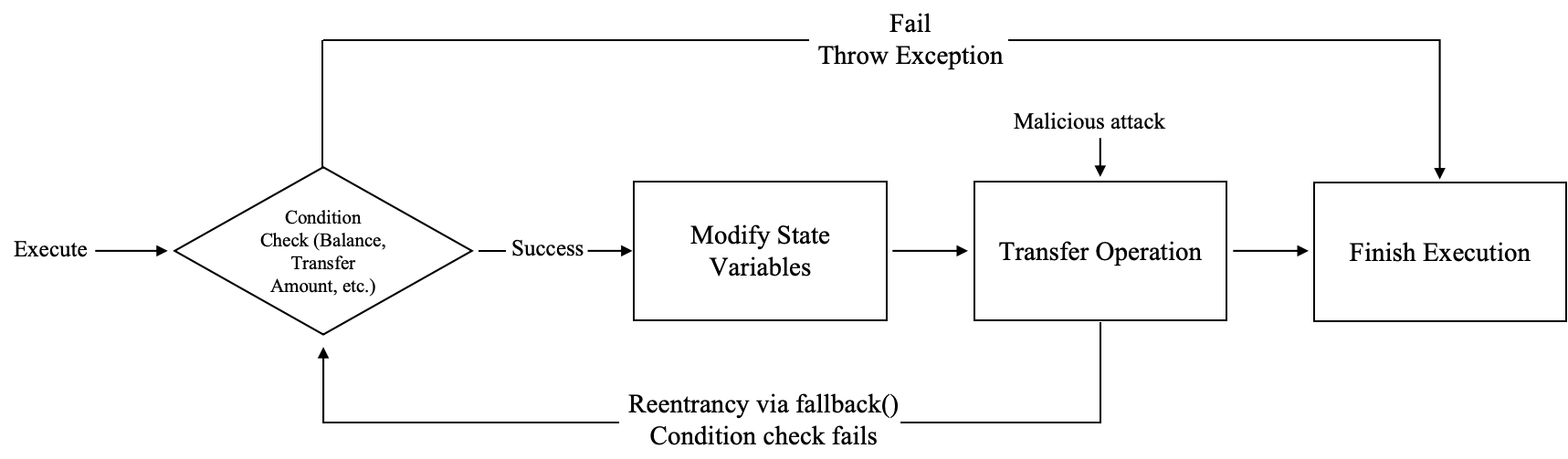}
        \caption{Security Pattern Workflow}
        \label{fig:3.3}
    \end{figure}

    Based on the reentrancy attack principle, the vulnerability arises because the contract performs Ether transfers before updating internal state variables. This allows attackers to bypass conditional checks and execute recursive reentrant calls. To mitigate this, the contract should update critical state variables before any external interaction, ensuring that subsequent calls are blocked by the updated state. The workflows of vulnerability patterns and security patterns are shown in Figures \ref{fig:3.2} and \ref{fig:3.3}, respectively.

   \textbf{(2) Introducing State Locks}
   
\begin{figure}
        \centering
        \includegraphics[width=1\linewidth]{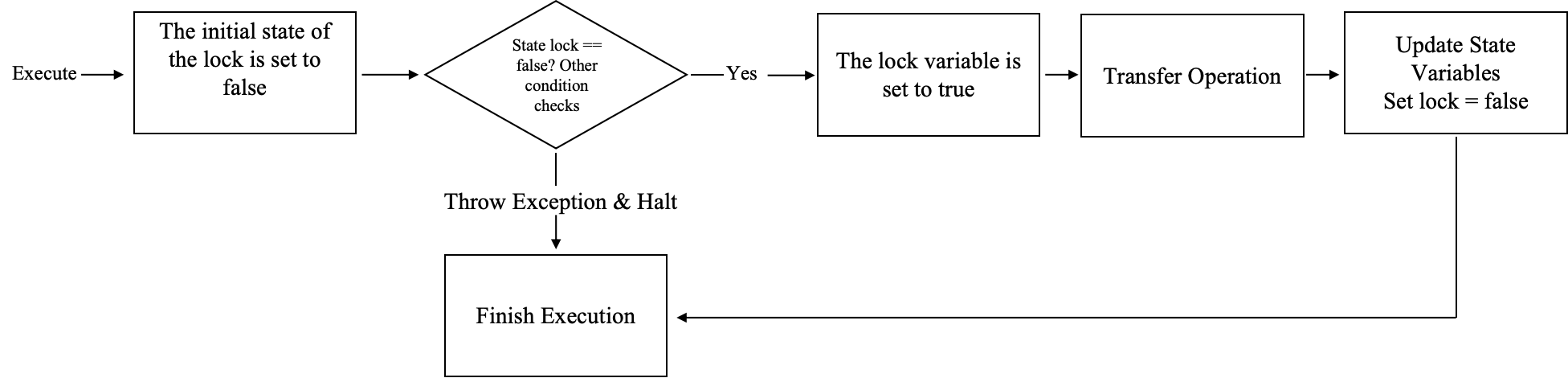}
        \caption{State Locks Workflow}
        \label{fig:3.4}
        \vspace{-1em}
    \end{figure}

    A lock mechanism can be employed to track whether the contract is in the middle of execution. A Boolean state variable is introduced to serve as a lock, which is set before executing sensitive logic and reset only after the function completes. This prevents any reentrant call from being successfully executed while the contract is locked, thus mitigating reentrancy. The workflow of state locks is shown in Figure \ref{fig:3.4}

    \textbf{(3) Using \texttt{transfer()} Instead of \texttt{call()}, or Limiting \texttt{gas}}
    
    Ethereum employs a gas mechanism to limit computational resource consumption, prevent abuse, and mitigate DoS attacks. Each transaction specifies a \texttt{gasLimit}, which defines the maximum gas it can consume. When a transaction is executed, the EVM checks the remaining gas before each operation. If the remaining gas is insufficient, the transaction reverts and consumes all the provided gas as a transaction fee.

\begin{figure}[htp]
        \centering
        \includegraphics[width=1\linewidth]{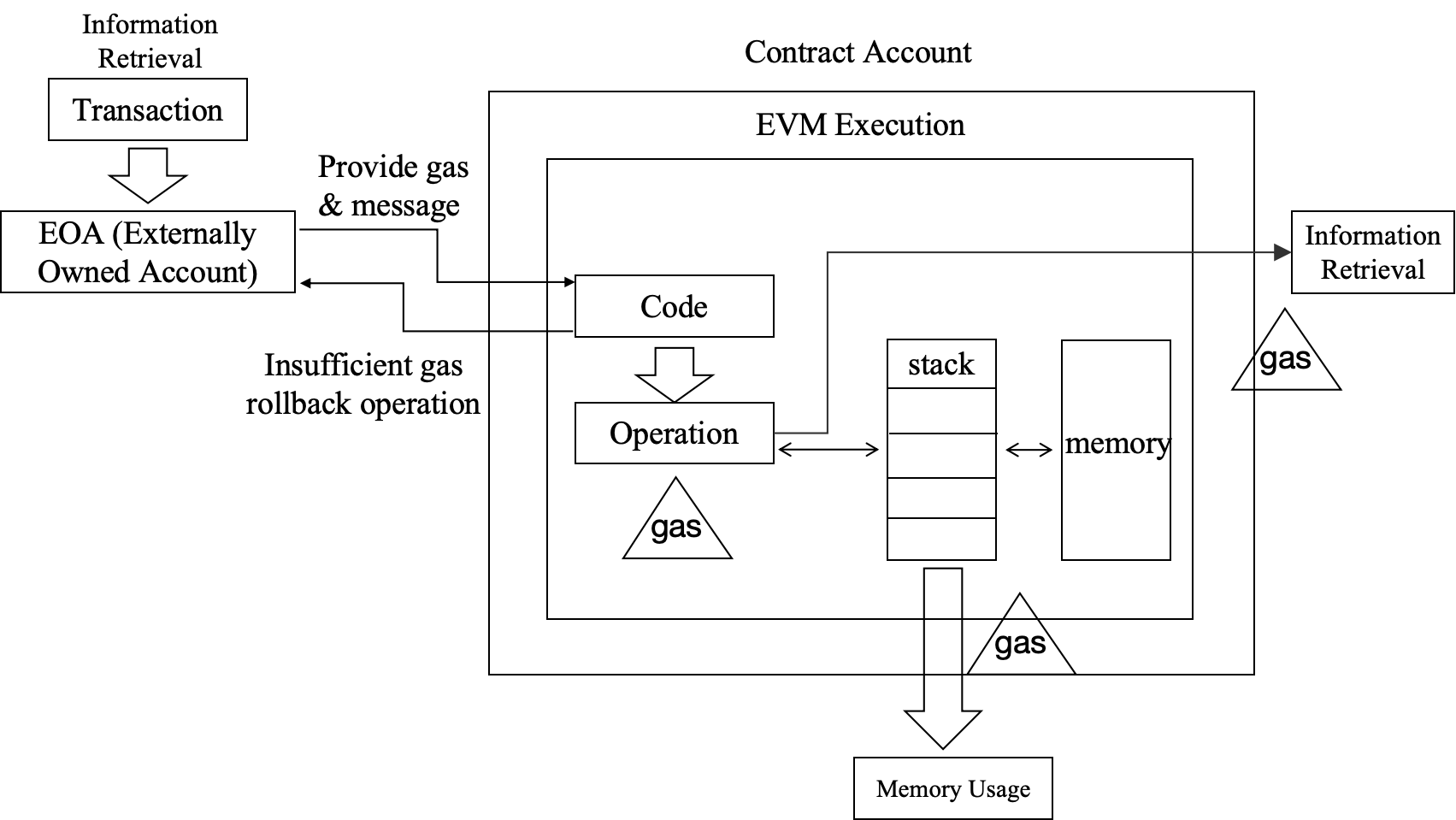}
        \caption{The Principle of the Gas Mechanism}
        \label{fig:3.5}
    \end{figure}

    The \texttt{call()} function forwards all remaining gas by default, which makes it susceptible to reentrancy attacks as attackers can execute complex fallback logic. In contrast, the \texttt{transfer()} function forwards only 2300 gas—insufficient for state-changing operations—thus significantly reducing the risk of reentrancy. Alternatively, developers may explicitly set the gas limit in \texttt{call()} to minimize exposure. The principle of the gas mechanism is illustrated in Figure \ref{fig:3.5}.

    \textbf{(4) Shifting Ether Control to the Receiver (Withdrawal Pattern)}
    
\begin{figure}[htp]
    \centering
    \includegraphics[width=0.8\linewidth]{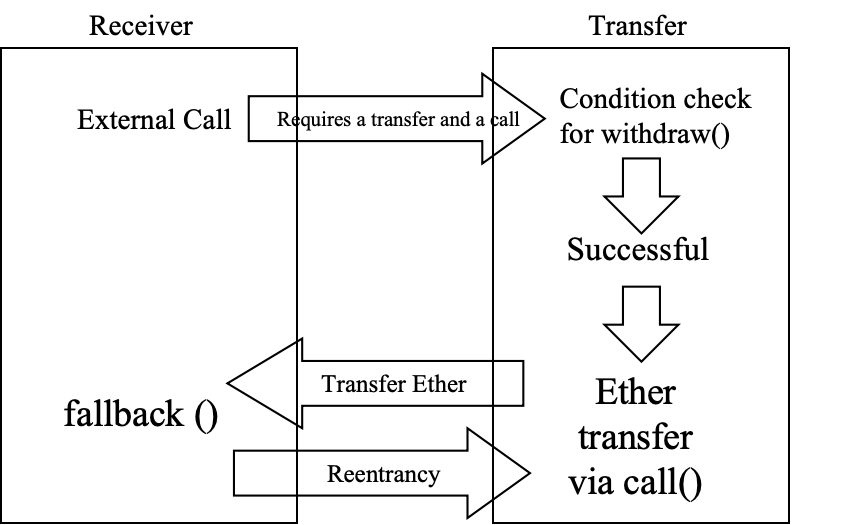}
    \caption{Transfer Pattern Workflow}
    \label{fig:3.6}
\end{figure}

\begin{figure}[htp]
    \centering
    \includegraphics[width=0.8\linewidth]{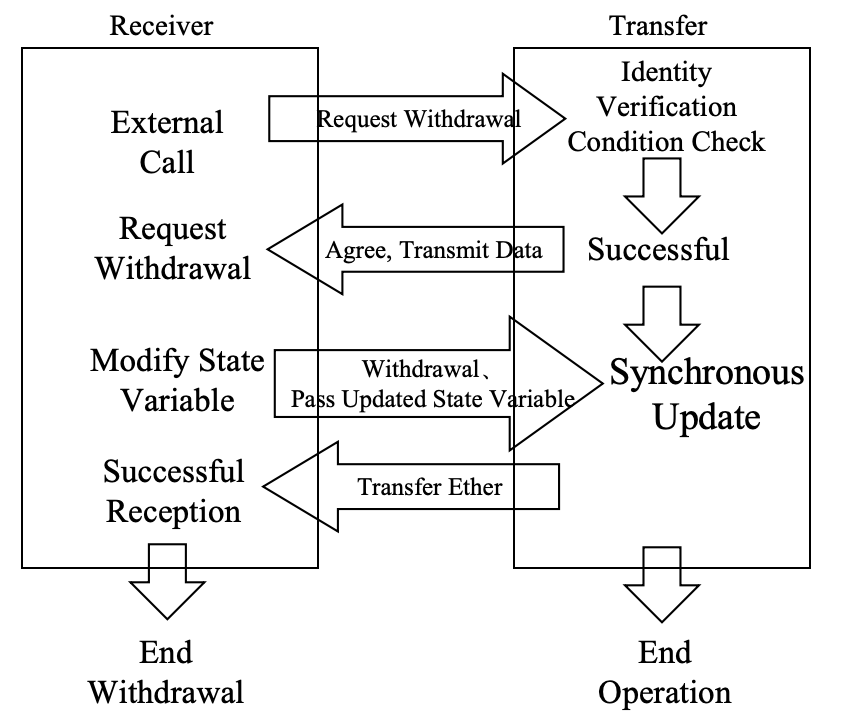}
    \vspace{-1em}
    \caption{Receiver Pattern Workflow}
    \label{fig:3.7}
    \vspace{-1em}
\end{figure}

    In the attack scenario, the vulnerable contract actively transferred Ether to the malicious contract using \texttt{call()}, which triggered the fallback function and enabled reentrancy. A more secure pattern is to shift control of the Ether transfer to the receiver contract, allowing the recipient to withdraw funds instead. This reduces logical coupling between state updates and Ether transfers, thereby minimizing the attack surface. The workflows of the transfer pattern and the receiver pattern are shown in Figures \ref{fig:3.6} and \ref{fig:3.7}, respectively.

    However, this pattern assumes that the recipient is trustworthy and that each withdrawal is legitimate. The callee contract must verify the identity and validity of the transaction initiator. Furthermore, this pattern increases the complexity and operational cost on the receiver side, especially when the number of participants is large. The added complexity may lead to higher gas consumption, which is not ideal for blockchain environments where efficiency is critical.

\subsection{Integer Overflow Vulnerability}

\subsubsection{Vulnerability Principle}
In Solidity, supported integer types range from uint8 to uint256 or int8 to int256, with step sizes in multiples of 8. A variable of type \texttt{uintX} (where $8 \leq X \leq 256$) represents unsigned integers ranging from 0 to $2^X - 1$. In the EVM, integers are fixed-size and unsigned, which introduces the risk of arithmetic overflows and underflows due to the limited numerical range of integer variables.
Three primary types of overflow exist: multiplication, addition, and subtraction overflow. They can lead to either an integer overflow (exceeding the maximum representable value) or underflow (falling below the minimum representable value).

For example: 

\begin{lstlisting}[language=Solidity]
uint8(*2) = uint8(0);
uint8(255 + 1) = uint8(0);
uint8(0 - 1) = uint8(255);
\end{lstlisting}

These vulnerabilities occur when inputs are not properly validated, and neither the Solidity compiler nor the EVM performs automatic overflow checks \cite{sun2021mutation}. Malicious users can exploit this behavior by crafting input data that causes arithmetic operations to wrap around, resulting in incorrect logic execution.

\subsubsection{Code Analysis} 

We analyze the following two code examples:

\textbf{(1) BEC Overflow Case}

\begin{figure}
    \centering
    \includegraphics[width=1\linewidth]{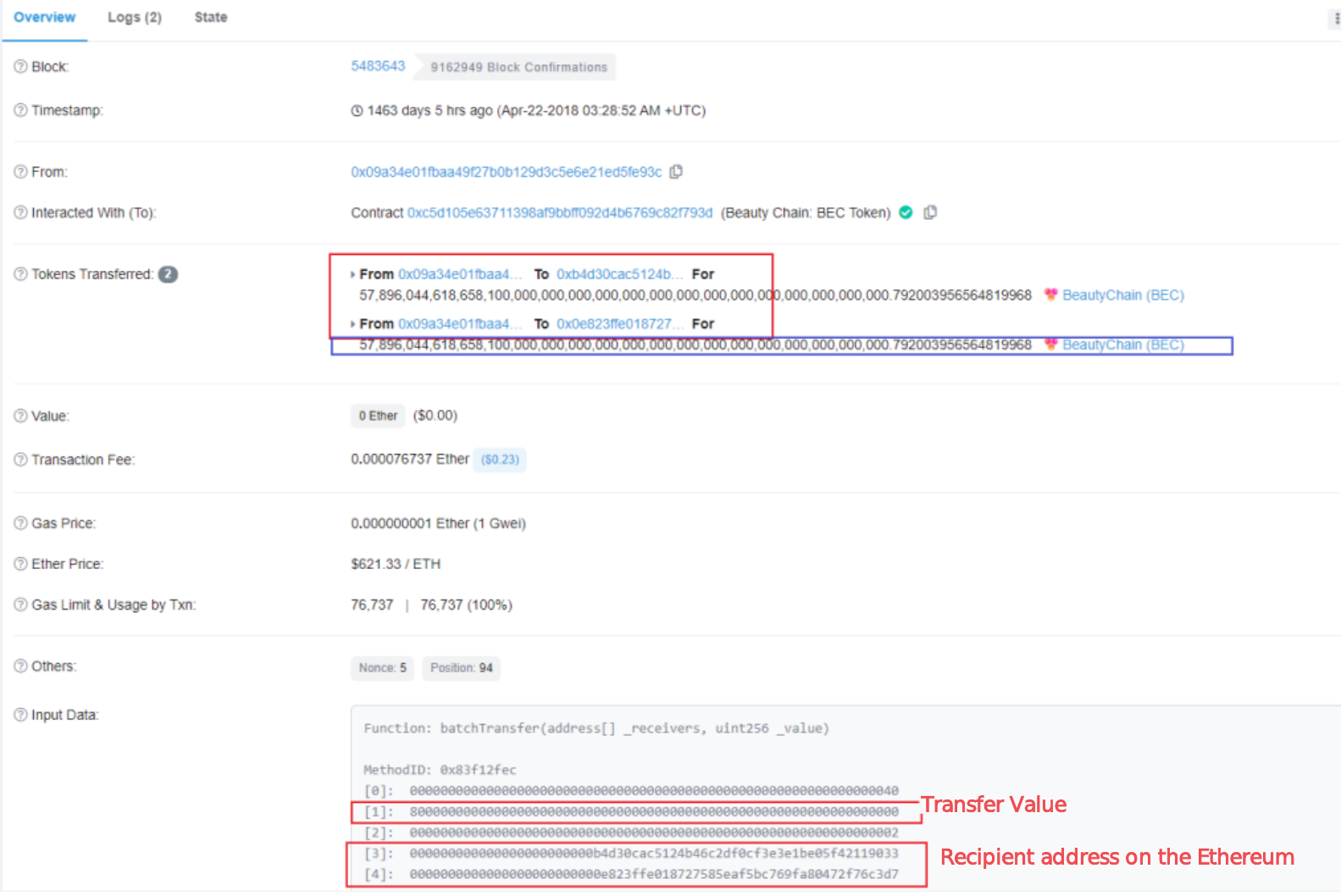}
    \vspace{-1em}
    \caption{Transaction Flow Diagram of the BEC Attack Event}
    \label{fig:BEC Attack}
     \vspace{-1em}
\end{figure}

Figure~\ref{fig:BEC Attack} illustrates the transaction trace of a real-world BEC overflow attack, in which a large amount of Ether was transferred to two recipient addresses.

The key code fragment vulnerable to overflow is shown in the Figure \ref{fig:Key Code}. The attacker controls the recipient address and the \texttt{\_value} parameter.

\begin{lstlisting}[language=Solidity]
amount = uint256(cnt) * _value;
\end{lstlisting}

By constructing an input where \texttt{cnt = 2} and \texttt{\_value = 2$^{255}$}, the multiplication overflows and results in \texttt{amount = 0}. As a result, the \texttt{require()} check is bypassed, and token transfers proceed without reducing the sender's token balance. This leads to the creation of a large number of tokens out of thin air, causing severe economic consequences, while the contract remains unaware of the exploit.

\begin{figure}[htp]
    \centering
    \includegraphics[width=0.8\linewidth]{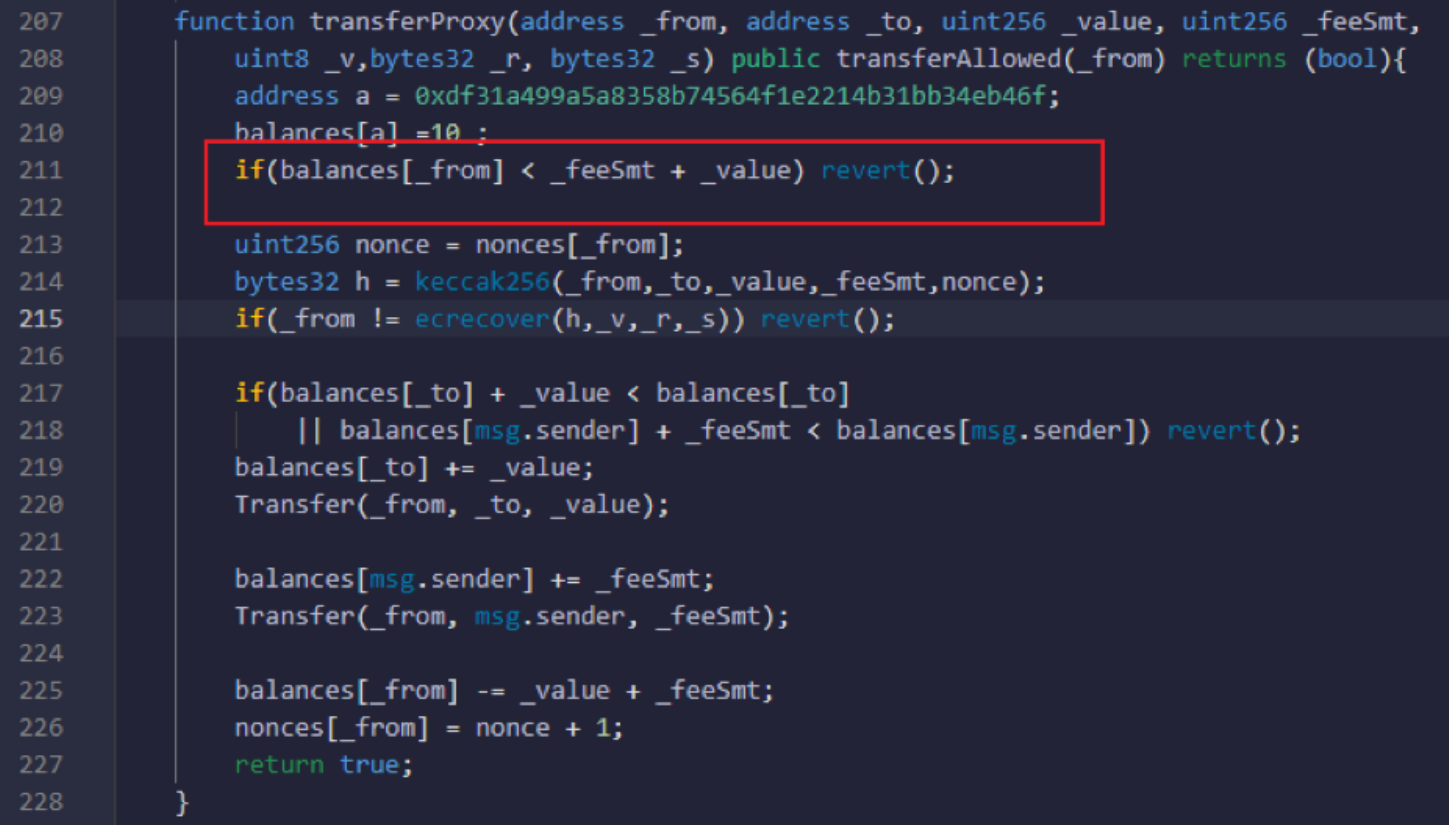}
    \caption{Key Code of the SMT Integer Overflow Attack}
    \label{fig:Key Code of the SMT}
    \vspace{-1em}
\end{figure}

\begin{figure}[H]
    \centering
    \includegraphics[width=0.95\linewidth]{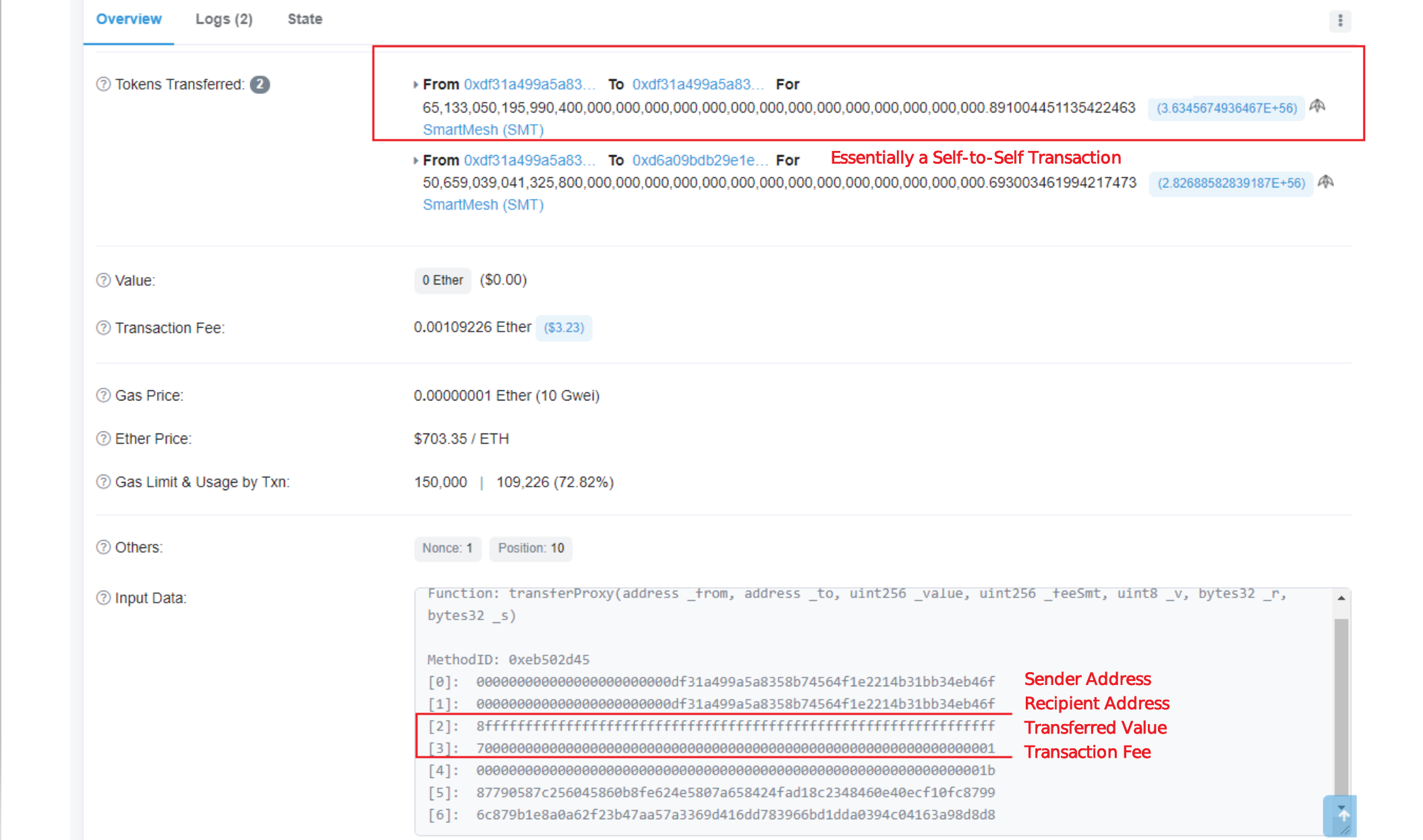}
        \vspace{-1em}
    \caption{Transaction Flow Diagram of the SMT Integer Overflow Attack}
    \label{fig:SMT Integer Overflow Attack}
    \vspace{-1em}
\end{figure}

\textbf{(2) SMT Overflow Case}
Figure~\ref{fig:SMT Integer Overflow Attack} and Figure \ref{fig:Key Code of the SMT} illustrate the transaction process and key code of the SMT overflow vulnerability.

In this case, the function checks whether the sum of \texttt{\_value + \_feeSmt} exceeds the sender’s balance. However, an attacker can construct inputs such that:
\begin{lstlisting}[language=Solidity]
_value + _feeSmt = 0x8fffffffff...ffff + 0x70000000...0001 == 2^256
\end{lstlisting}

\begin{figure}[htp]
    \centering
    \includegraphics[width=0.8\linewidth]{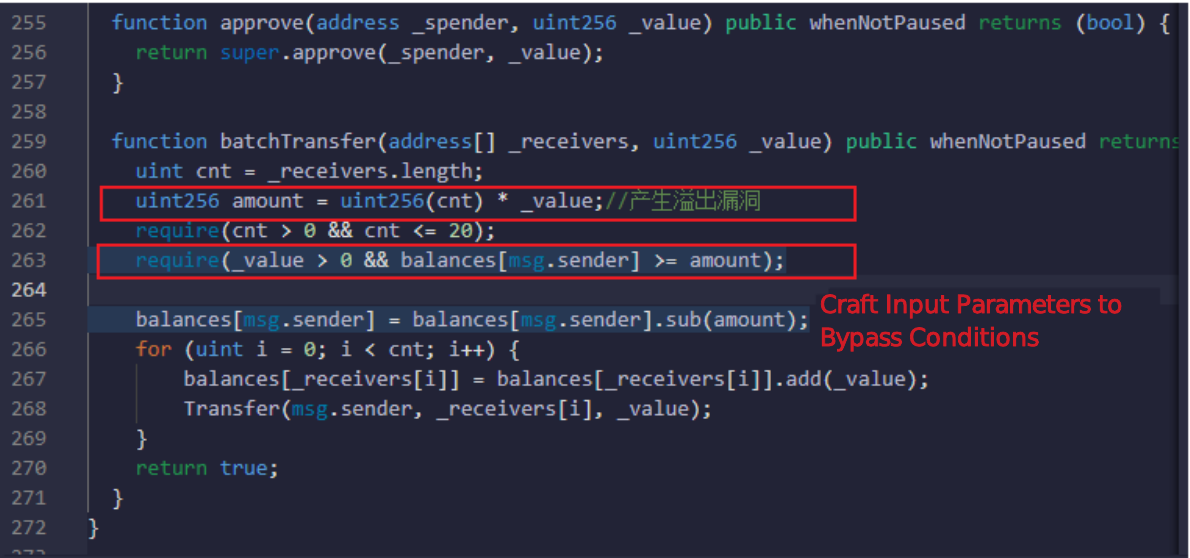}
    \caption{Key Code of BEC Integer Overflow Attack}
    \label{fig:Key Code}
    \vspace{-1em}
\end{figure}

In this way, an overflow can be triggered. As a result, the computed value becomes 0, bypassing the if condition check. Execution then proceeds with the digital signature verification and token distribution operations. The integer overflow attack is thus successfully executed, causing no decrease in the token balance of the sender address and a significant increase in the balance of the recipient address. In this case, since the sender and recipient addresses are the same, the account balance increases substantially without any actual transfer of tokens.

\subsubsection{Impacts and Consequences}
Integer overflow vulnerabilities allow attackers to bypass conditional logic and obtain excessive amounts of Ether or tokens. In severe cases, this may result in the infinite minting of tokens or forged token transfers. Such exploits violate the intended logic of smart contracts. If exploited by a contract administrator, this type of vulnerability could lead to abuse and catastrophic damage to the system.

\subsubsection{Security Pattern Design}
\begin{figure}[H]
    \centering
    \includegraphics[width=0.9\linewidth]{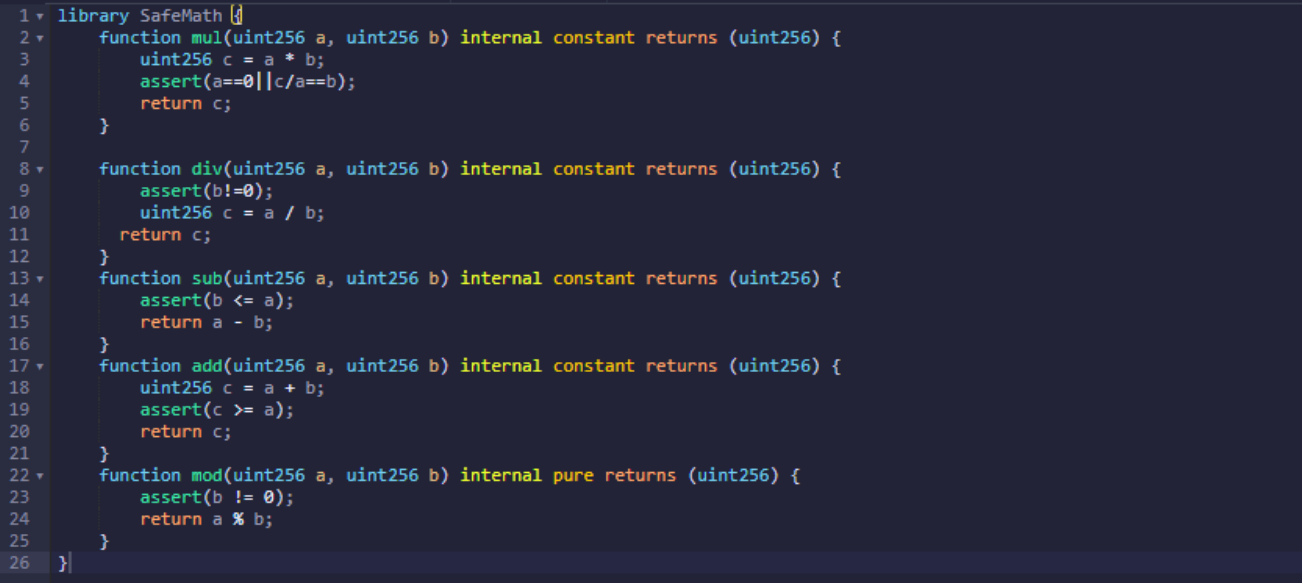}
    \vspace{-1em}
    \caption{Code of SafeMath}
    \label{fig:safemath}
    \vspace{-1em}
\end{figure}

To mitigate overflow vulnerabilities, arithmetic operations should be accompanied by rigorous pre- and post-condition validations. The SafeMath library \cite{openzeppelin}, provided by OpenZeppelin, wraps arithmetic logic with assertions to prevent overflows. Figure \ref{fig:safemath} is an example of SafeMath usage.

The arithmetic logic issues arising from integer overflow are effectively validated and handled in SafeMath. The  \texttt{assert()} is used for verification, ensuring that the computation satisfies the specified conditions. If the conditions are not met, the execution is halted, and an error is triggered. By utilizing the arithmetic interfaces encapsulated in the SafeMath library, integer overflow vulnerabilities are effectively mitigated.

\section{Experimental Reproduction}
\subsection{Experimental Environment and Tools}
\begin{itemize}
    \item \textbf{Operating System:} Windows 10
    \item \textbf{Tools Used:} Remix IDE (original and Chinese versions), Geth Ethereum client
    \item \textbf{Programming Languages:} Solidity (contract-oriented language for Ethereum), Go
\end{itemize}

\subsection{Experimental Results and Analysis}
\subsubsection{Reentrancy Vulnerability}

\paragraph{\textbf{(1) Vulnerable Mode}}

\subparagraph{(i) Contract Deployment and Balance Inquiry.}

\begin{figure}[H]
    \centering
    \includegraphics[width=0.9\linewidth]{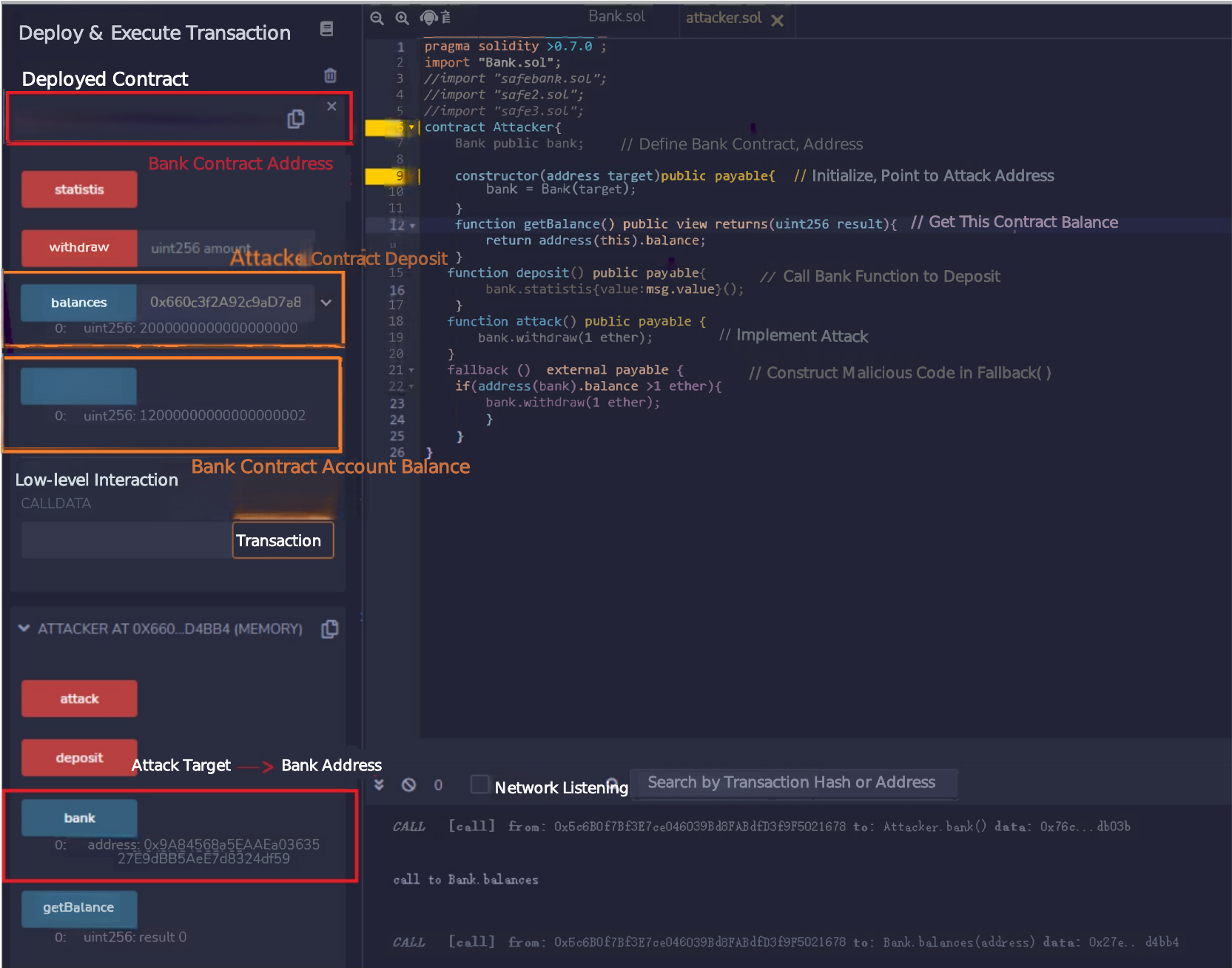}
    \caption{Deployment of Vulnerable Contract for Reentrancy Attack}
    \label{fig:4.1}
    \vspace{-1em}
\end{figure}

The process of contract deployment is shown in Figure \ref{fig:4.1}.

\subparagraph{(ii) Reentrancy Attack Execution.} 

\begin{figure}[H]
    \centering
    \includegraphics[width=1\linewidth]{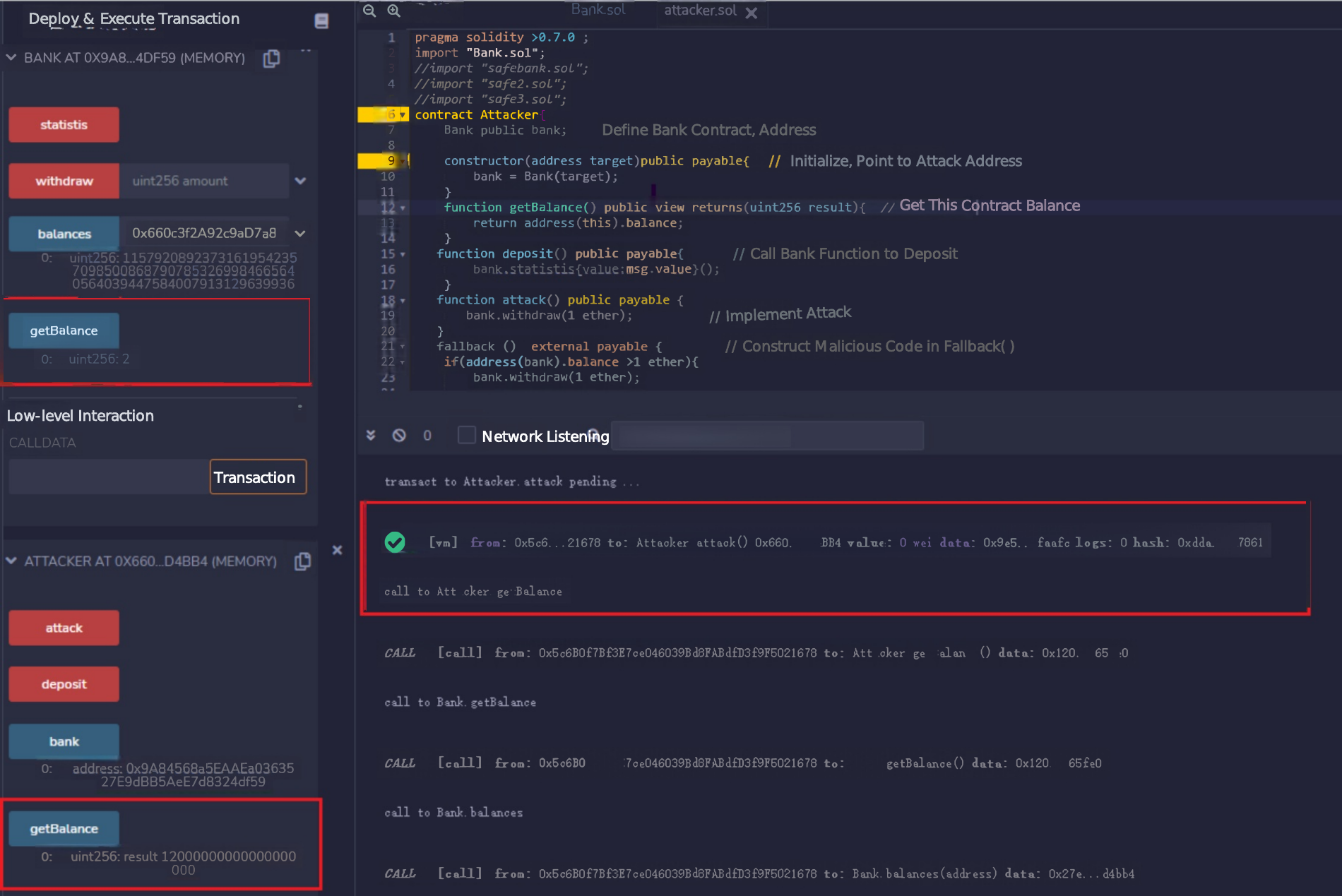}
    \caption{Execution of the reentrancy attack}
    \label{fig:4.2}
\end{figure}

The process of the execution of the reentrancy attack is shown in Figure \ref{fig:4.2}.

\subparagraph{(iii) Attack Results.}

\begin{figure}[H]
    \centering
    \begin{minipage}[b]{0.45\linewidth}
        \centering
        \includegraphics[width=\linewidth]{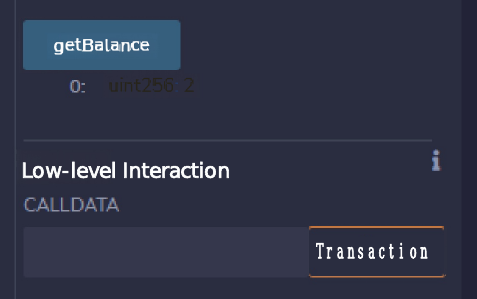}
        \caption{Balance of Bank Contract After The Attack}
        \label{fig:4.3}
    \end{minipage}
    \hspace{0.00005\linewidth} 
    \begin{minipage}[b]{0.5\linewidth}
        \centering
        \includegraphics[width=\linewidth]{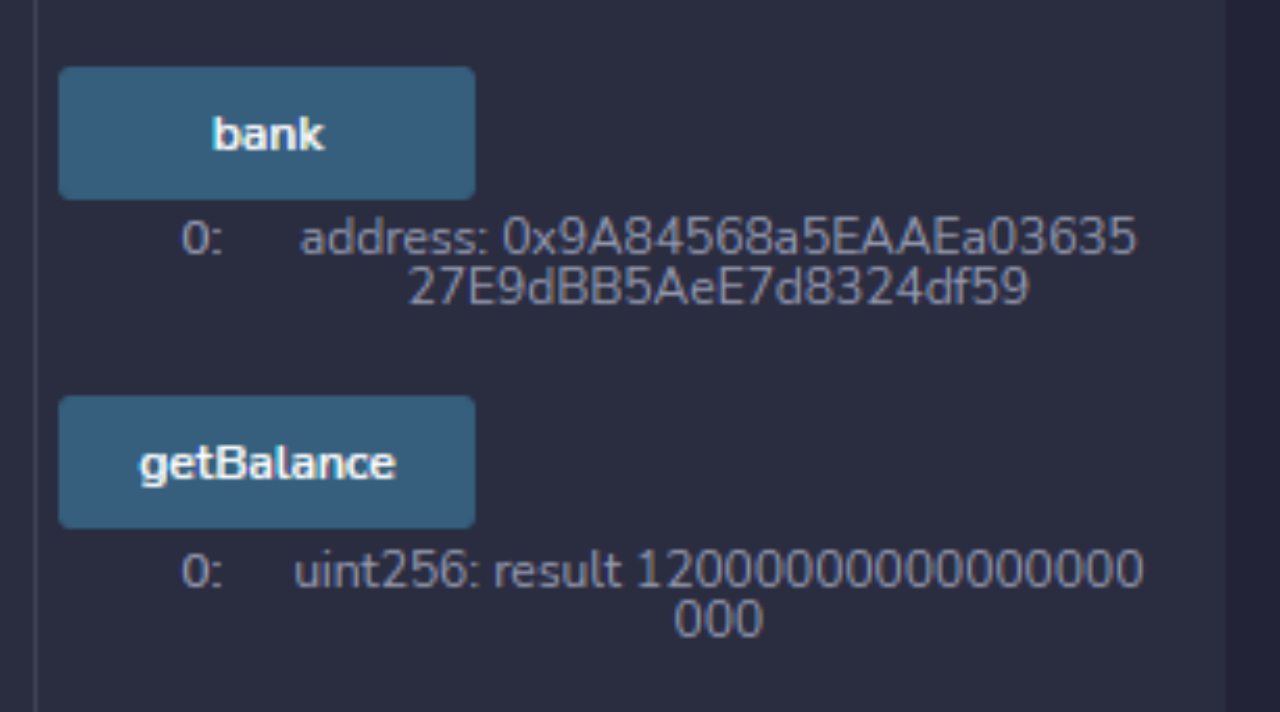}
        \caption{Balance of Attacker After The Attack}
        \label{fig:4.3l}
    \end{minipage}
\end{figure}

The balance of the Bank contract after the attack is shown in Figure \ref{fig:4.3}. Besides, the balance of the attacker is shown in Figure \ref{fig:4.3l}.



   \subparagraph{ (iv) Transaction Process.}

\begin{figure}[H]
        \centering
        \includegraphics[width=0.5\linewidth]{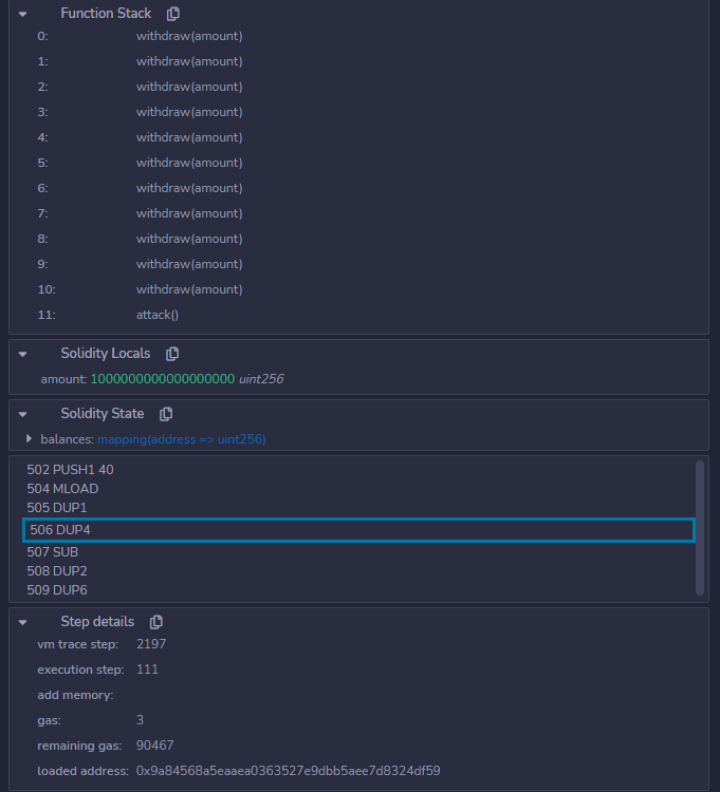}
        \includegraphics[width=0.49\linewidth]{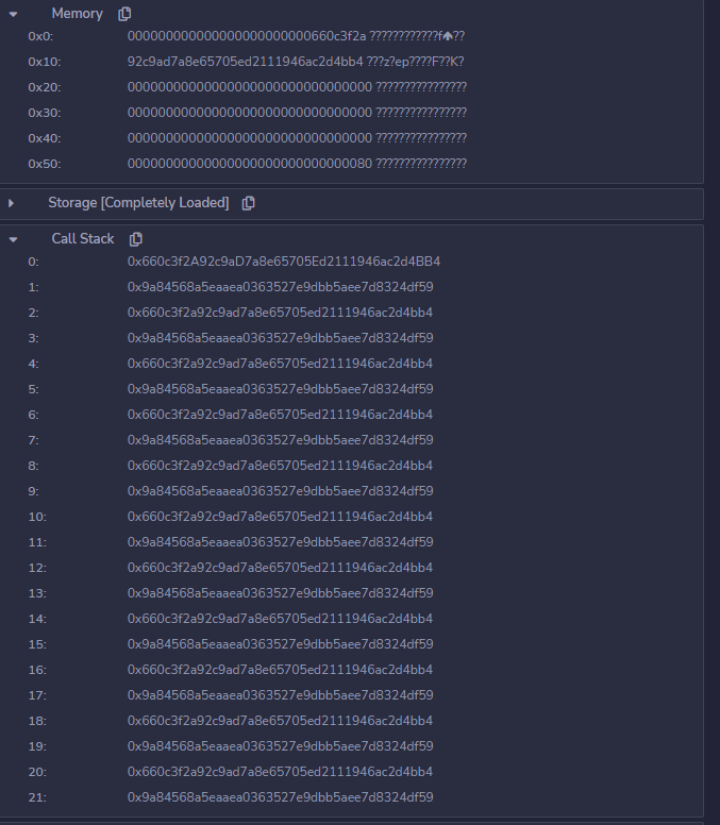}
        \caption{The Transaction Processes of Attack}
        \label{fig:4.4.12}
    \end{figure}

Figure \ref{fig:4.4.12} illustrates the function call process during the procedure, highlighting multiple reentrancy attacks on the withdraw transfer function, which ultimately lead to the successful theft of tokens.


\paragraph{\textbf{(2) Secure Modes}}
\subparagraph{(i) Checks-Effects-Interactions Pattern.}
The code uses the Checks-Effects-Interactions (CEI) pattern, in which state variables are updated before performing any external calls.

\begin{figure}[H]
    \centering
    \includegraphics[width=0.9\linewidth]{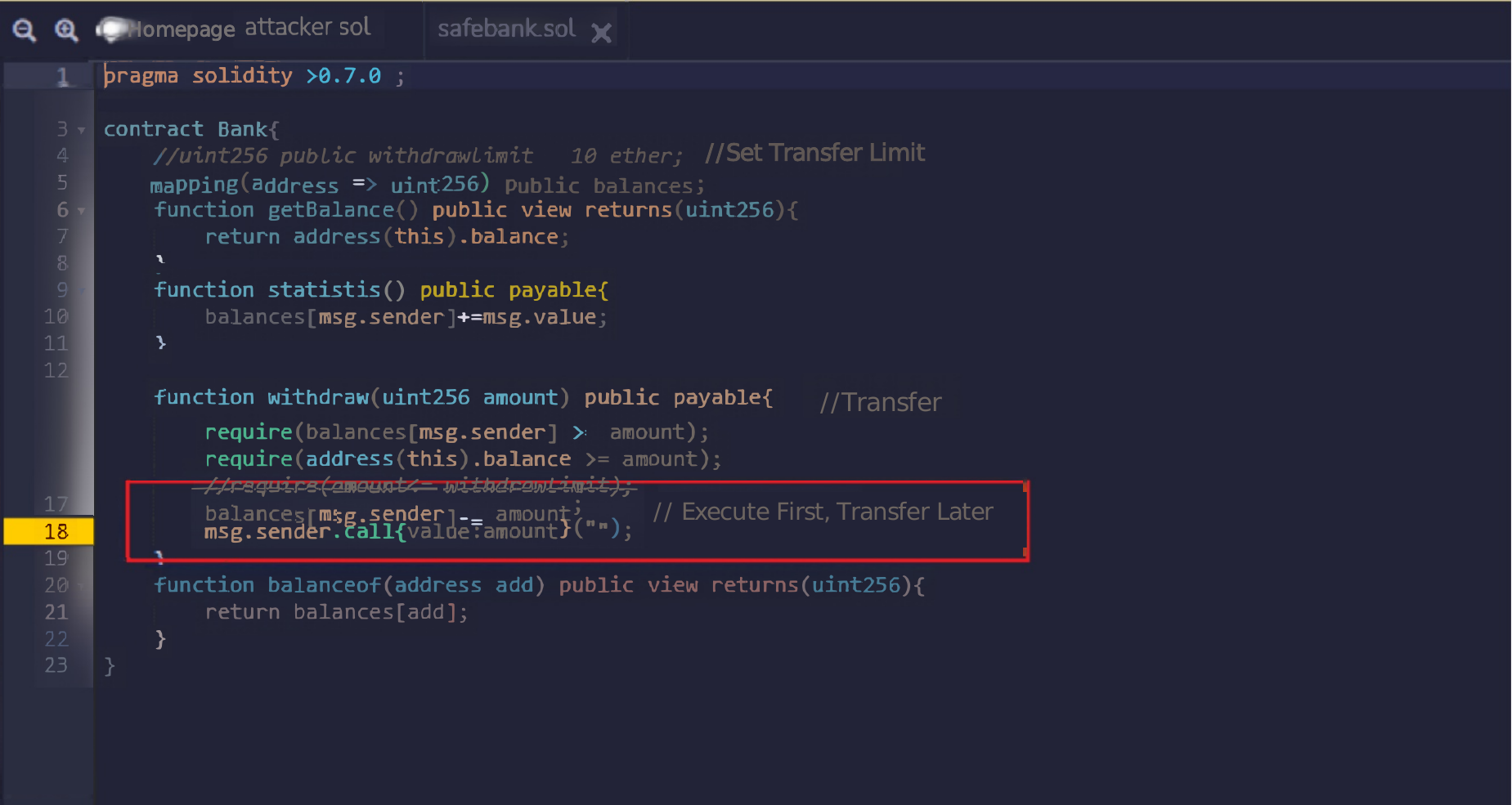}
    \caption{Secure Contract Using CEI Pattern}
    \label{fig:4.5}
    \vspace{-1em}
\end{figure}

Figure \ref{fig:4.5} shows the code in the secure mode. The red-marked parts highlight the differences between the secure contract and the vulnerable one.

\textbf{Attack Outcome:} 

\begin{figure}[H]
    \centering
    \includegraphics[width=0.9\linewidth]{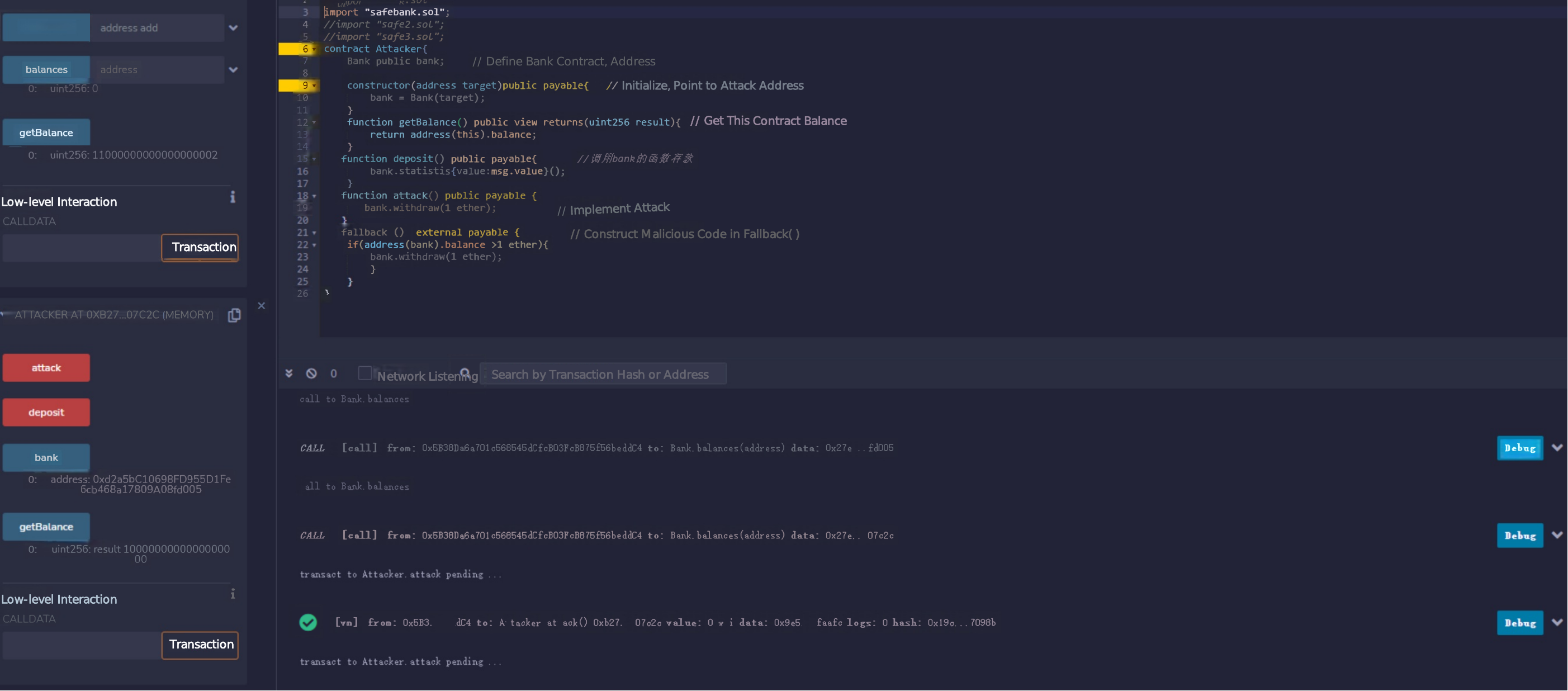}
    \caption{Attack Execution Results}
    \label{fig:4.6}
    \vspace{-1em}
\end{figure}

As shown in Figure \ref{fig:4.6}, even when the attacker has 2 ethers in the contract, the internal balance becomes 0 after one withdrawal, not 1 ether as expected.

\begin{figure}[H]
    \centering
    \includegraphics[width=0.5\linewidth]{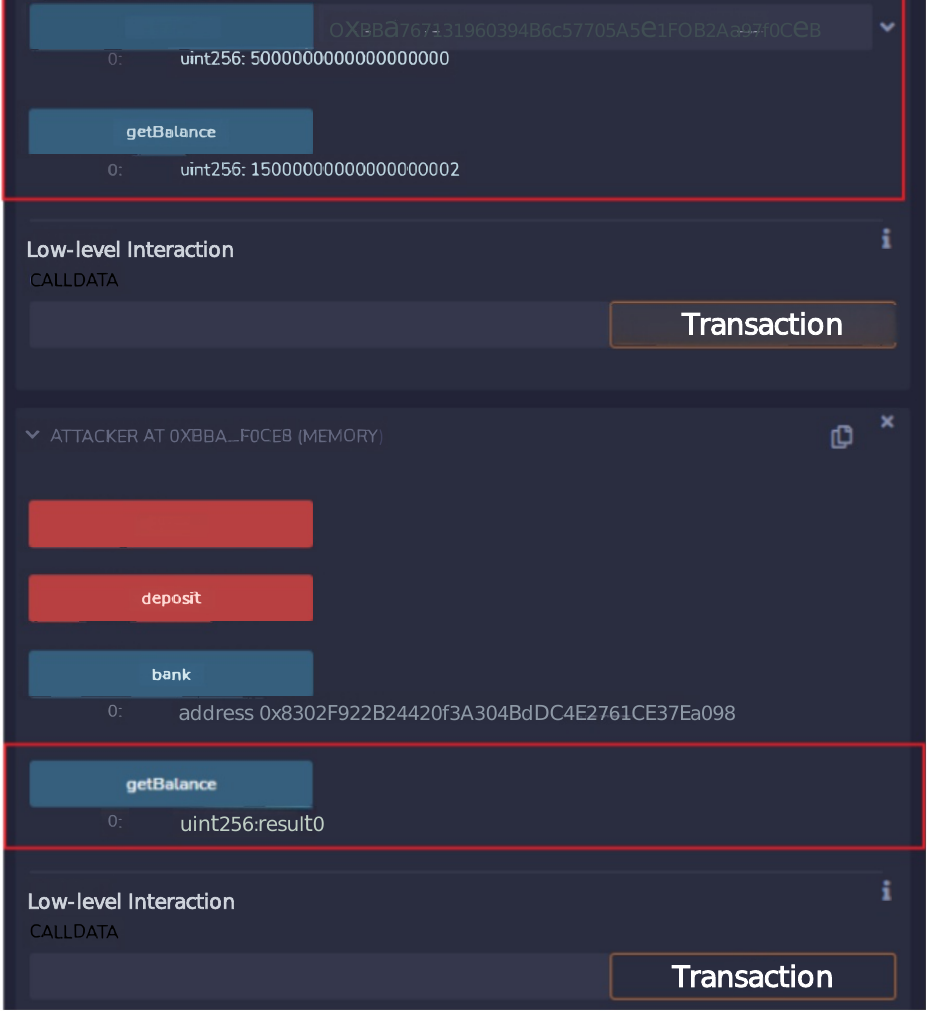}
    \includegraphics[width=0.49\linewidth]{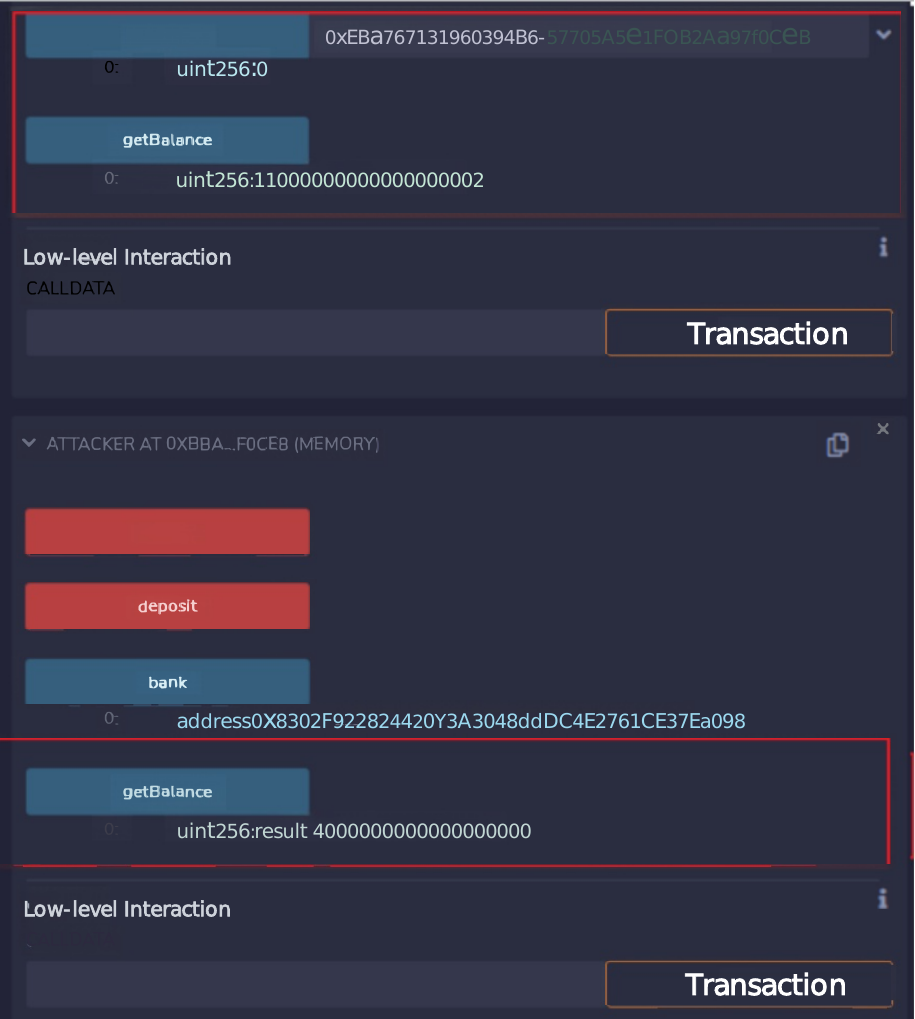}
    \vspace{-0.5em}
    \caption{Initial and Final States of Attack With 5 Ether Deposit}
    \label{fig:4.78}
    \vspace{-1em}
\end{figure}

In the experimental setup, the Attacker has a deposit of 5 Ether in the Bank contract. The initial and resulting states are shown in Figure \ref{fig:4.78}. Although the intended transfer amount is 1 Ether, the contract performs a transfer based on a manipulated deposit value of - 1 Ether.

\begin{figure}[H]
    \centering
    \includegraphics[width=0.7\linewidth]{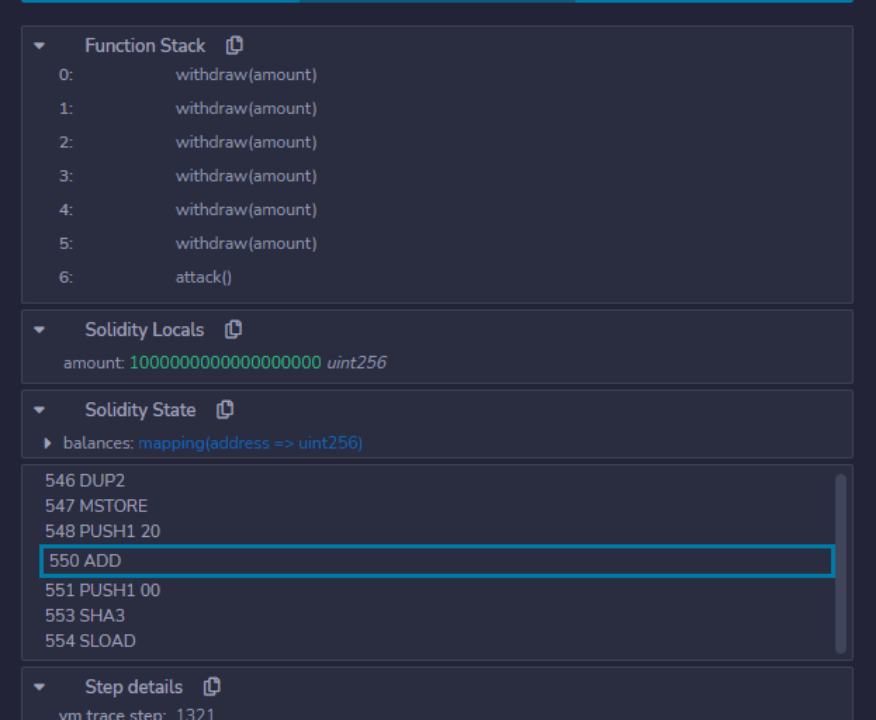}
    \caption{Execution Trace Showing Fallback Reentrancy}
    \label{fig:4.9}
    \vspace{-1em}
\end{figure}

\textbf{Analysis:}  
As shown in Figure \ref{fig:4.9}, although the balance is updated before the transfer, a fallback function in the attacker can still perform reentrant calls as long as the contract holds enough ether.

\begin{figure}[H]
    \centering
    \includegraphics[width=0.9\linewidth]{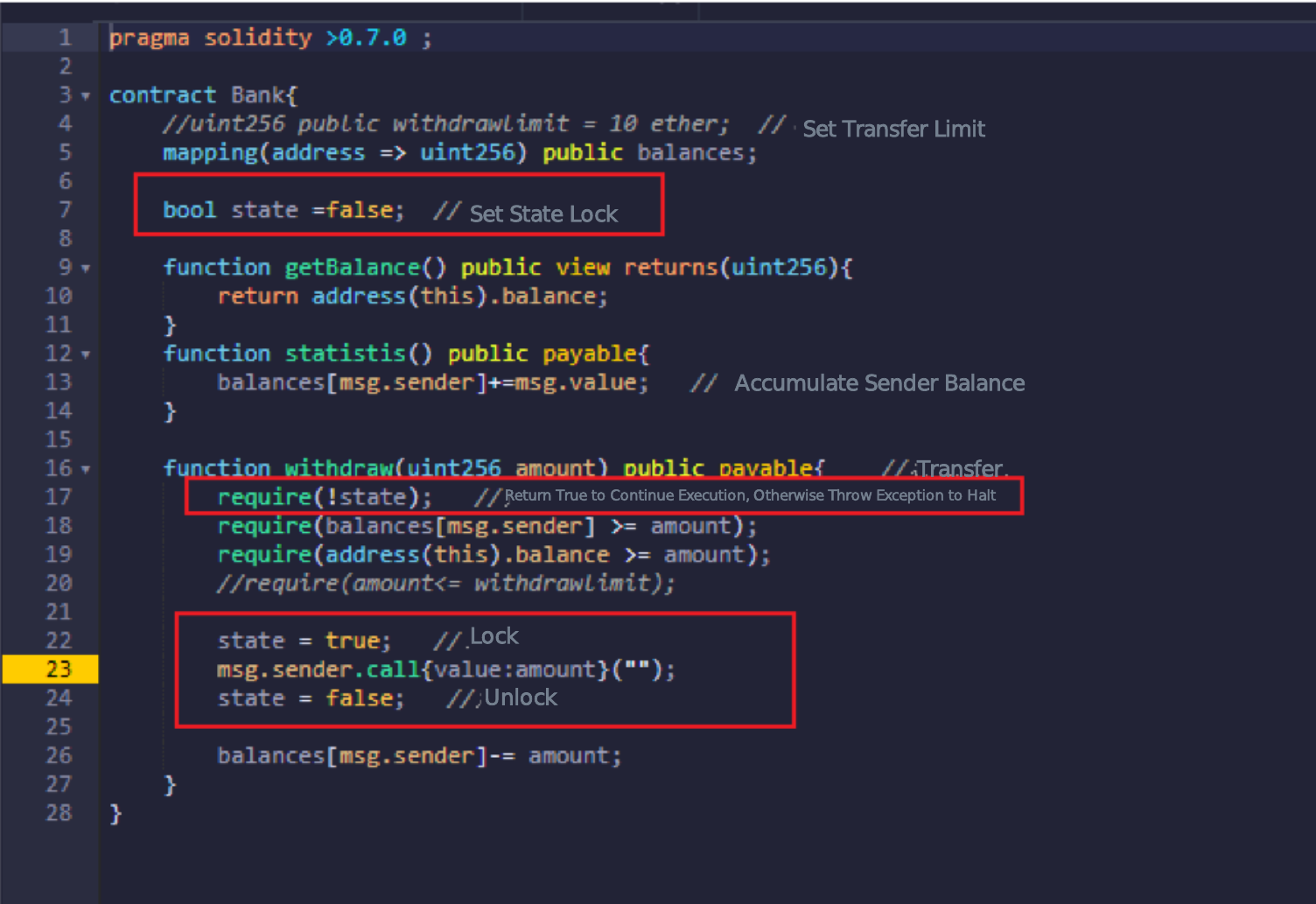}
    \caption{Secure contract using a state lock}
    \label{fig:4.10}
    \vspace{-1em}
\end{figure}

\subparagraph{(ii) Introducing a State Lock.}
As shown in Figure \ref{fig:4.10}, a state lock is introduced to prevent concurrent access.

\textbf{Execution Results:}

\begin{figure}[H]
    \centering
    \includegraphics[width=0.95\linewidth]{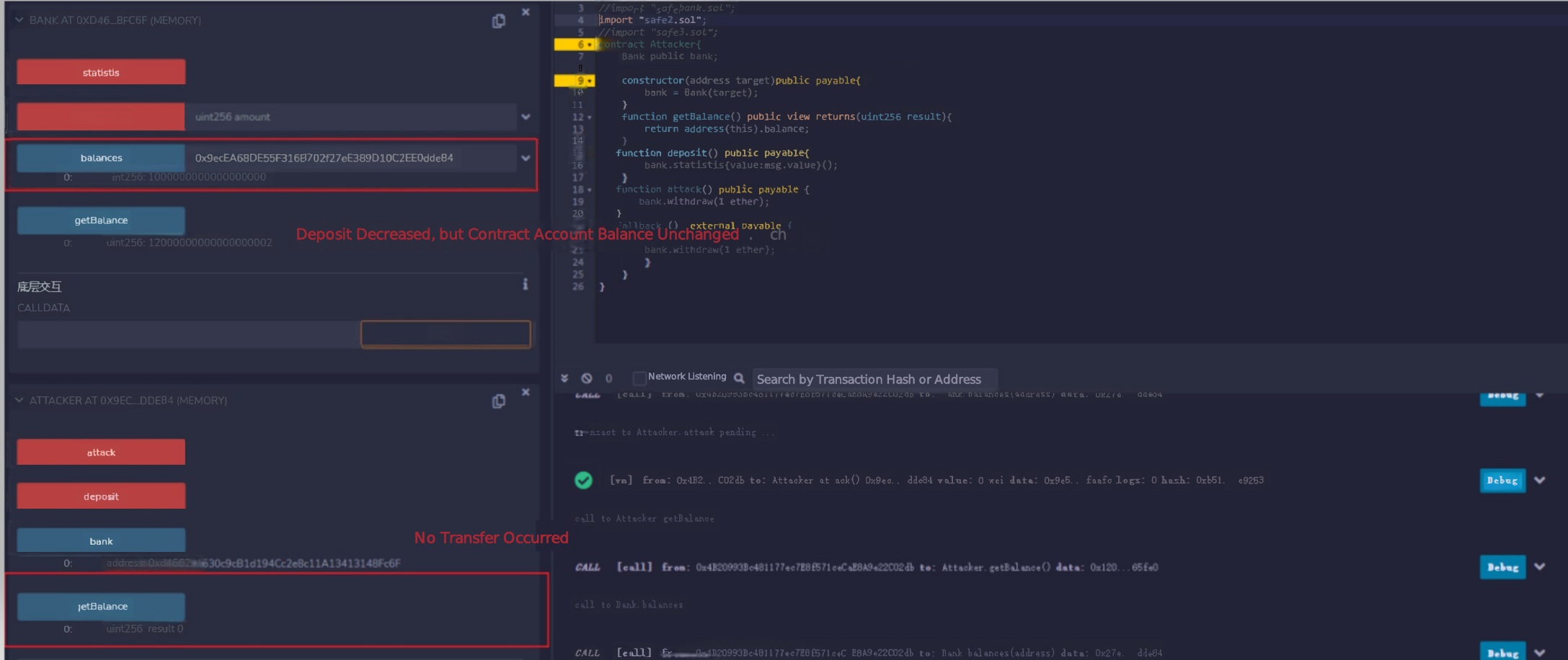}
    \caption{Final State of the Attack}
        \label{fig:4.12}
        \vspace{-1em}
\end{figure}

\begin{figure}[H]
    \centering
    \includegraphics[width=0.8\linewidth]{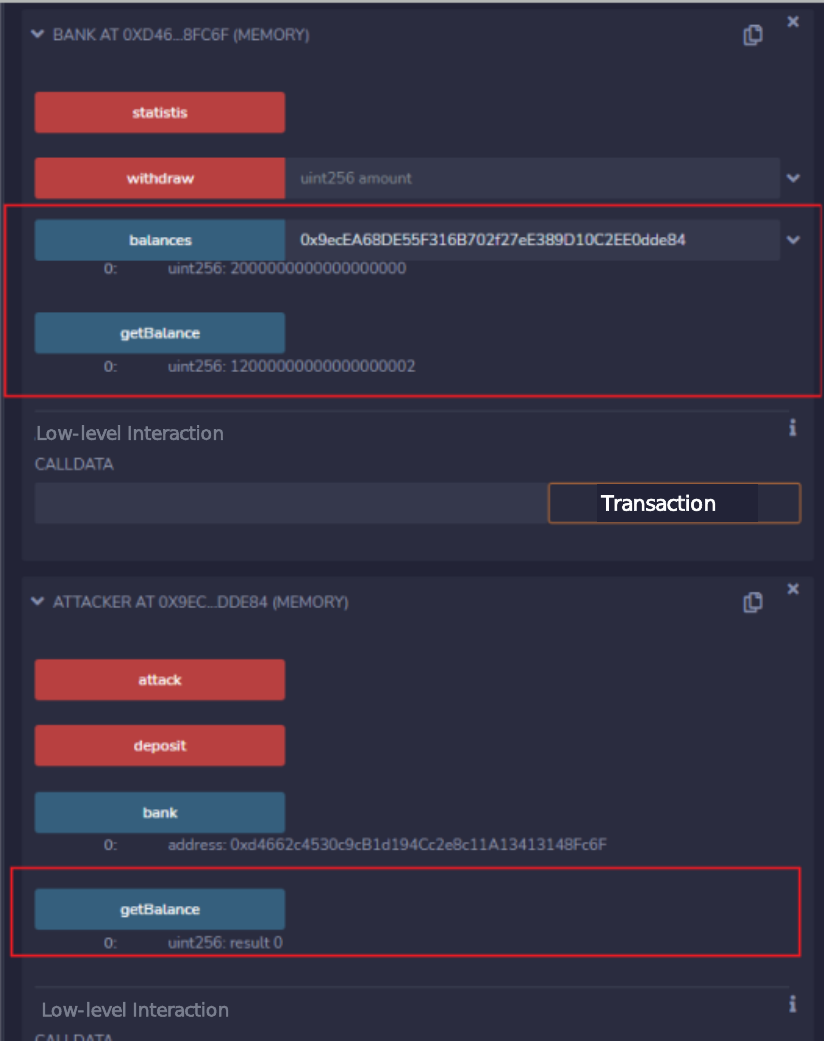}
    \caption{Initial State of the Attack}
     \label{fig:4.11}
\end{figure}

Figure \ref{fig:4.11} and Figure \ref{fig:4.12} show the initial state and the result after the attack.

\textbf{Analysis:}  
When a transfer is initiated, a state lock is set. However, during a reentrancy attack, the state lock has not yet been engaged, causing the conditional check to fail and the transfer to be aborted. Nevertheless, since the \texttt{call()} does not revert upon failure, execution continues, leading to the subsequent unlocking of the state and the reduction of the caller's deposit. This results in an inconsistency in the recorded deposit balance.

\subparagraph{(iii) Using \texttt{transfer()} Instead of \texttt{call()}.}

\begin{figure}[H]
    \centering
    \includegraphics[width=0.9\linewidth]{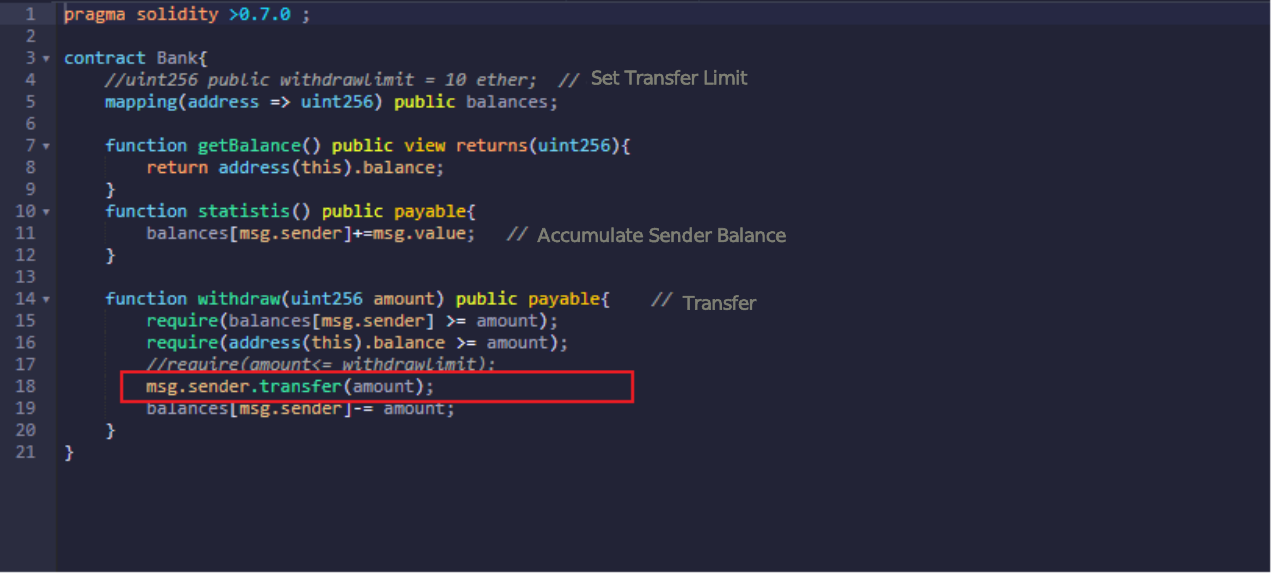}
    \caption{Secure contract Using \texttt{transfer()}}
    \label{fig:4.13}
    \vspace{-1em}
\end{figure}

Figure \ref{fig:4.13} presents the code for the secure version. The sections highlighted in red indicate the differences between the secure contract and the vulnerable contract.

\textbf{Execution Results:}

\begin{figure}[H]
    \centering
    \includegraphics[width=0.9\linewidth]{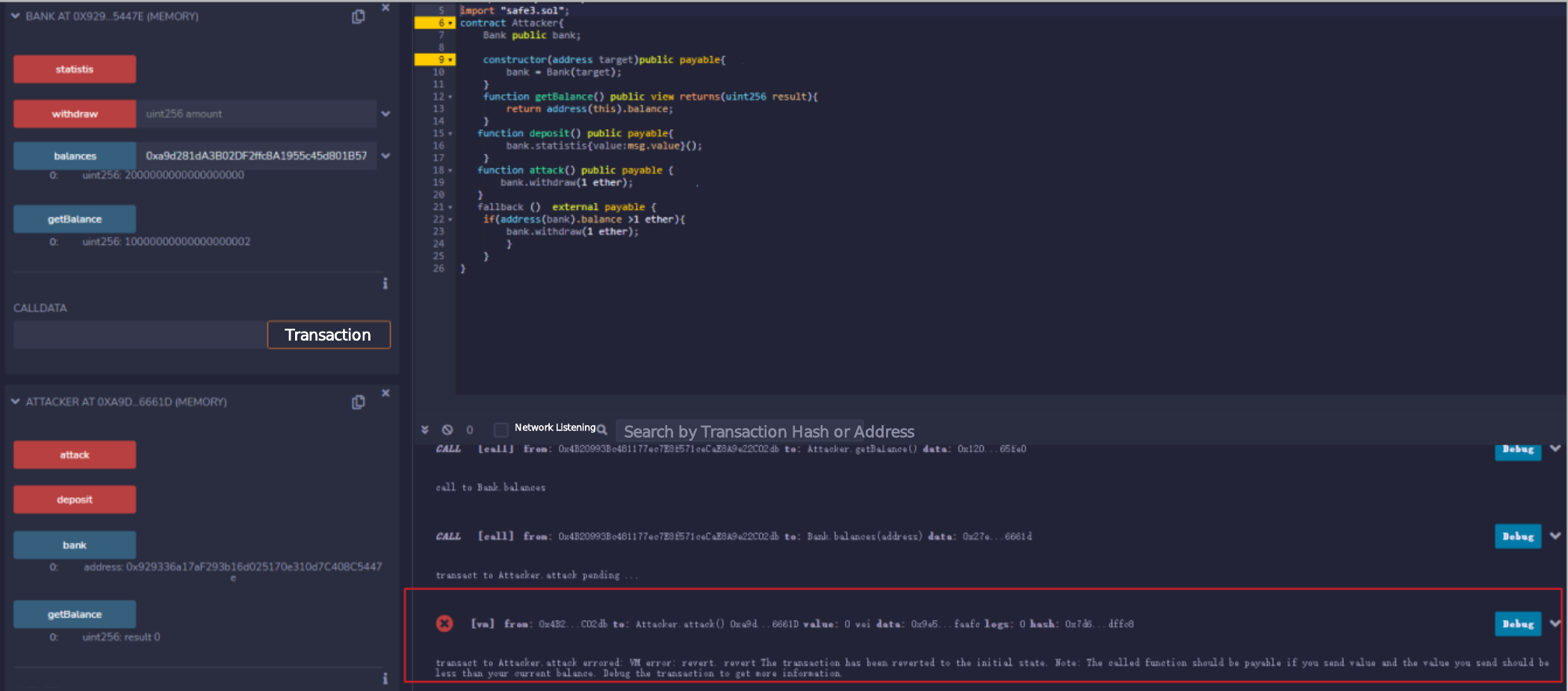}
    \caption{Attack Outcome by Using \texttt{transfer()}}
    \label{fig:4.14}
    \vspace{-1em}
\end{figure}

Figure \ref{fig:4.14} illustrates a failed transfer during the execution of the attack function. At this point, the Attacker contract contains a malicious fallback function.

\begin{figure}[H]
    \centering
    \includegraphics[width=0.5\linewidth]{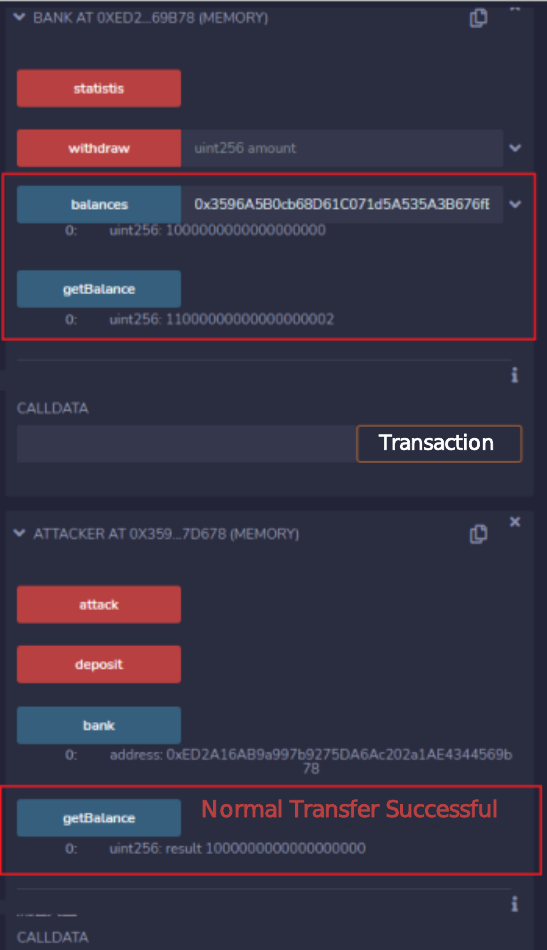}
    \caption{Correct Transfer Outcome by Using \texttt{transfer()}}
    \label{fig:4.15}
    \vspace{-1em}
\end{figure}

After removing the malicious fallback function, a normal transfer is performed, and the result is shown in Figure \ref{fig:4.15}. The transfer using \texttt{ transfer()} successfully completes the Ethereum transaction. When the call function in the contract contains a malicious operation that attempts to accept Ether transfers, an exception is thrown, halting execution.

\textbf{Analysis:}  
Due to the gas consumption required for blockchain operations, the \texttt{ transfer()} is only provided with 2300 gas, which is insufficient to support the Ether transfer operation in the fallback function. Additionally, the recursive calls caused by reentrancy attacks lead to significant gas consumption.

\subparagraph{(iv) Comparative Analysis of Secure Modes.}

\begin{table}[ht]
\centering
\caption{Comparative Analysis of Reentrancy Vulnerability Security Modes}
\begin{tabular}{p{3cm}|p{5.5cm}|p{6cm}}
\hline
\textbf{Security Mode}  & \textbf{Disadvantages} & \textbf{Reentrancy Limitation} \\ \hline
Check-Effect-Interaction    
& Allows limited reentrancy, potentially bypassing normal withdrawal conditions    
& Dependent on the caller's deposit and the contract's transfer restrictions, potentially exceeding normal withdrawal limits \\ \hline
State Lock    
& Causes inconsistency in deposits      
& Tied to normal withdrawal limits \\ \hline
Transfer() Function    
& Null
& High limitation, throws an exception and halts execution \\ \hline
Withdrawal Mode    
& Consumes significant computational resources and requires trust in the caller    
& Dependent on the trust level in the caller, logically reduces reentrancy risk \\ \hline
\end{tabular}
\end{table}

\subsubsection{Integer Overflow Vulnerability}
\paragraph{{\textbf{(1) Insecure Mode}}}

This section will reproduce the BEC event integer overflow vulnerability.
 
\subparagraph{(i). Contract deployment and initial state setting.}

\begin{figure}[H]
      \hspace*{0cm}
      \includegraphics[width=1.2\linewidth]{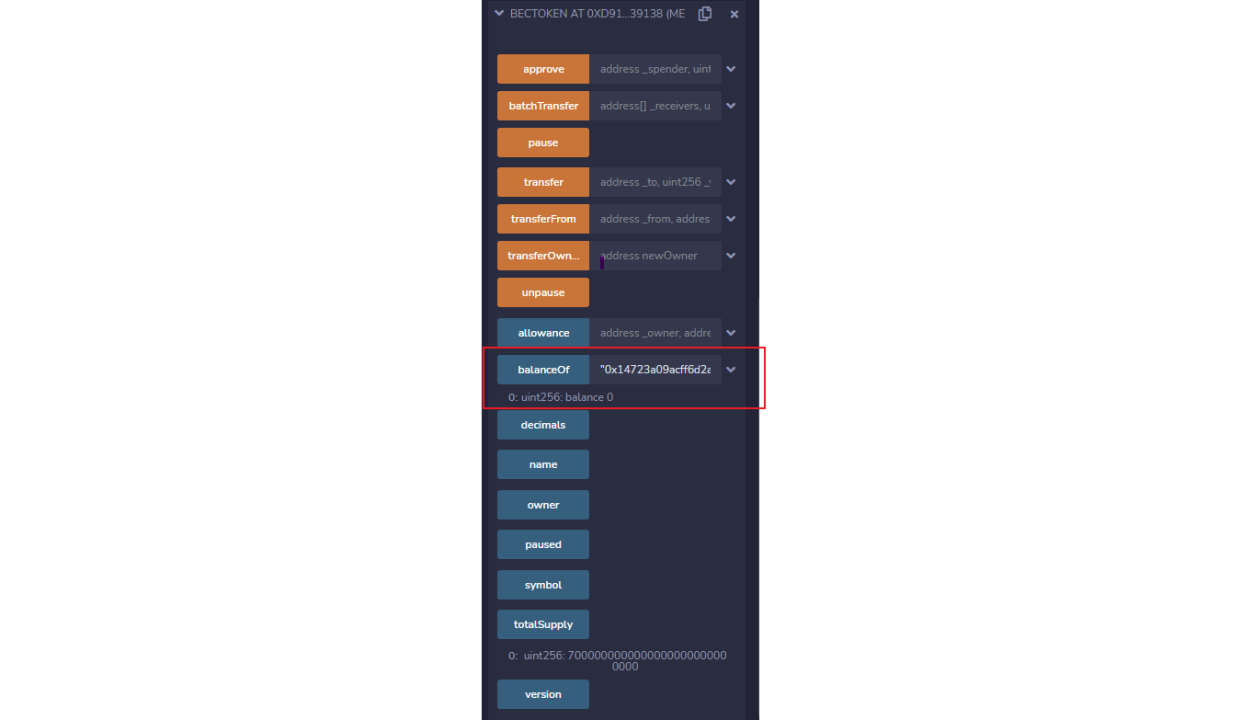}
      \caption{Initial State at Contract Deployment}
      \label{fig:4-16}
  \end{figure}

The initial state of the contract at deployment is shown in Figure \ref{fig:4-16}.

  \subparagraph{(ii). Attack Results.}

\begin{figure}[H]
      \centering
      \includegraphics[width=1\linewidth]{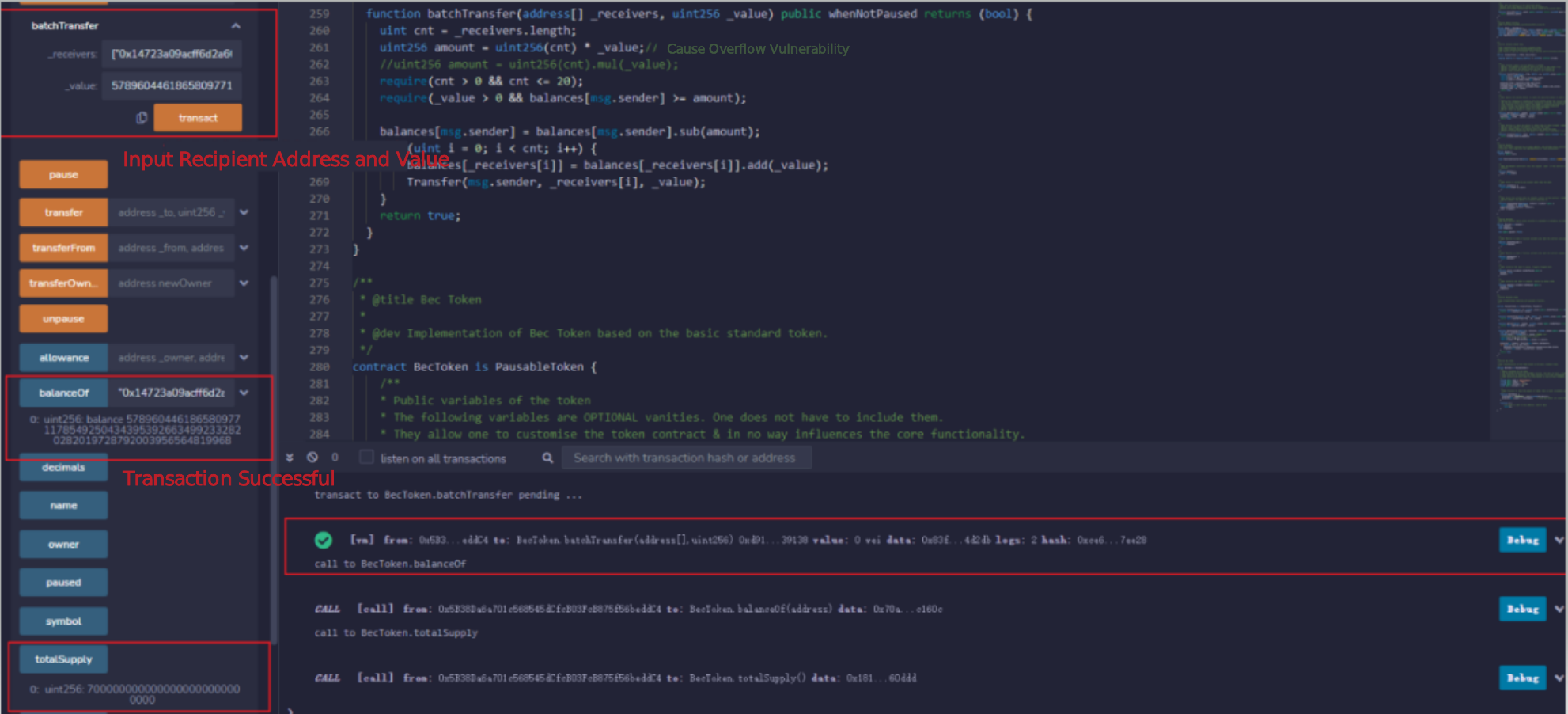}
      \caption{Attack Execution Result}
      \label{fig:4.17}
  \end{figure}

\begin{figure}[H]
      \centering
      \includegraphics[width=1\linewidth]{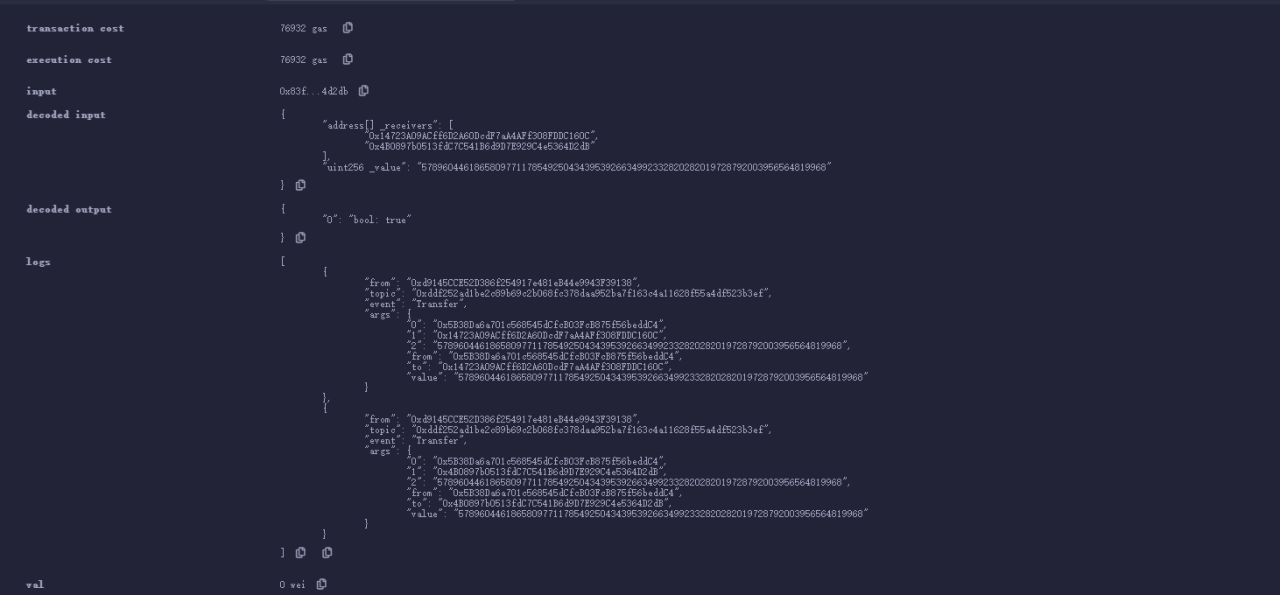}
      \caption{Attack Execution Transaction Trace}
      \label{fig:4.18}
  \end{figure}

The result of the executed attack is shown in Figure \ref{fig:4.17} and Figure \ref{fig:4.18}.

  \subparagraph{(iii). Result analysis.}

\begin{figure}[H]
      \centering
      \includegraphics[width=1\linewidth]{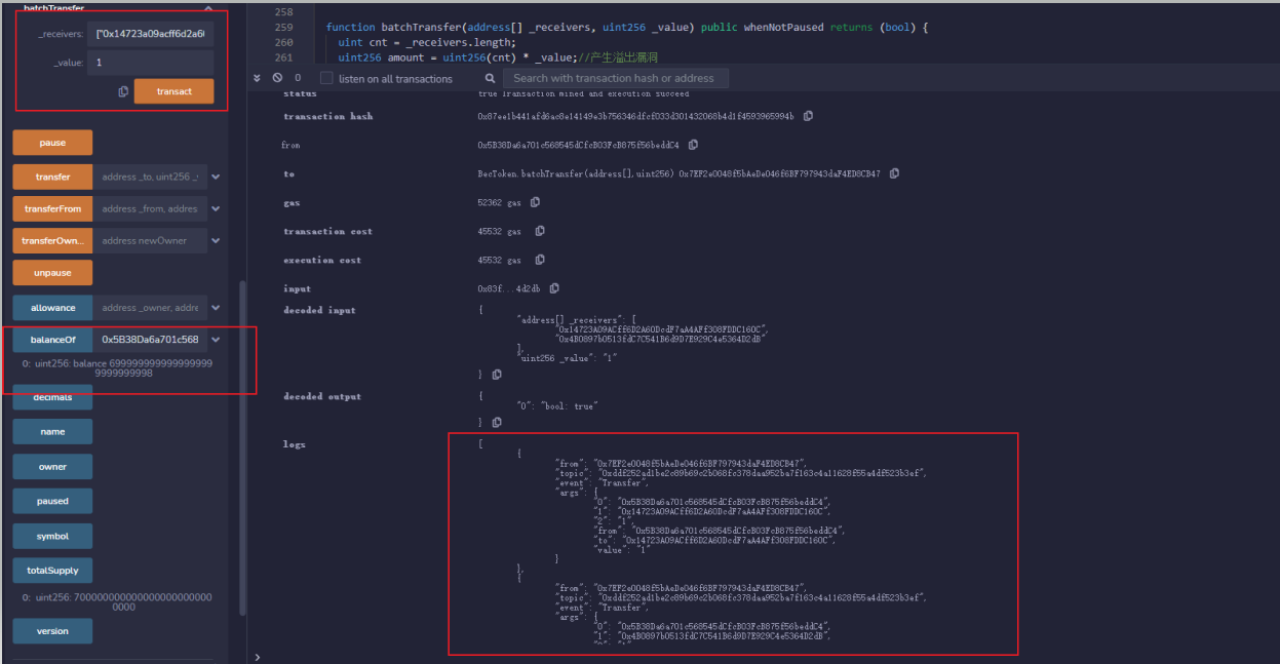}
      \caption{Correct Transaction Trace}
      \label{fig:4-19}
  \end{figure}

  Under normal operations, the successful transfer transaction is shown in Figure \ref{fig:4-19}.

\begin{figure}[htbp]
    \centering
    \begin{minipage}[b]{0.5\linewidth}
        \centering
        \includegraphics[width=\linewidth]{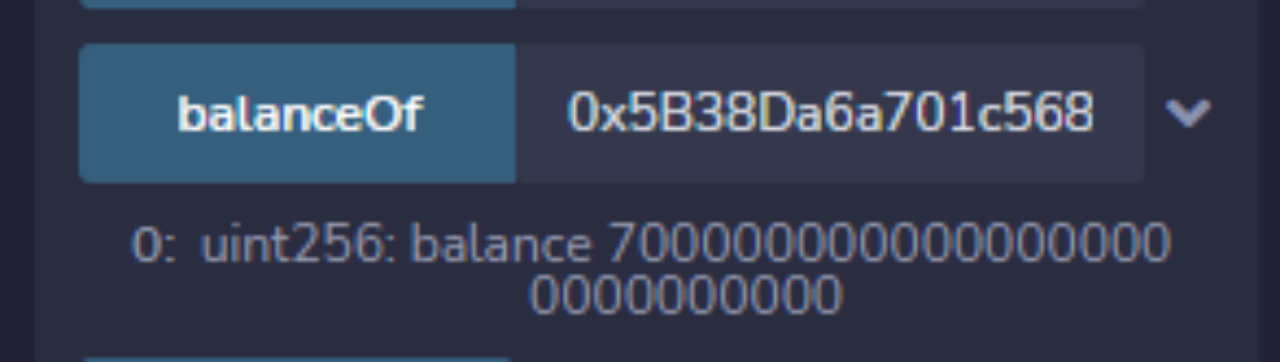}
        \caption{Abnormal Transaction Result (1)}
        \label{fig:4-20}
    \end{minipage}
    \hspace{0.0005\linewidth} 
    \begin{minipage}[b]{0.48\linewidth}
        \centering
        \includegraphics[width=\linewidth]{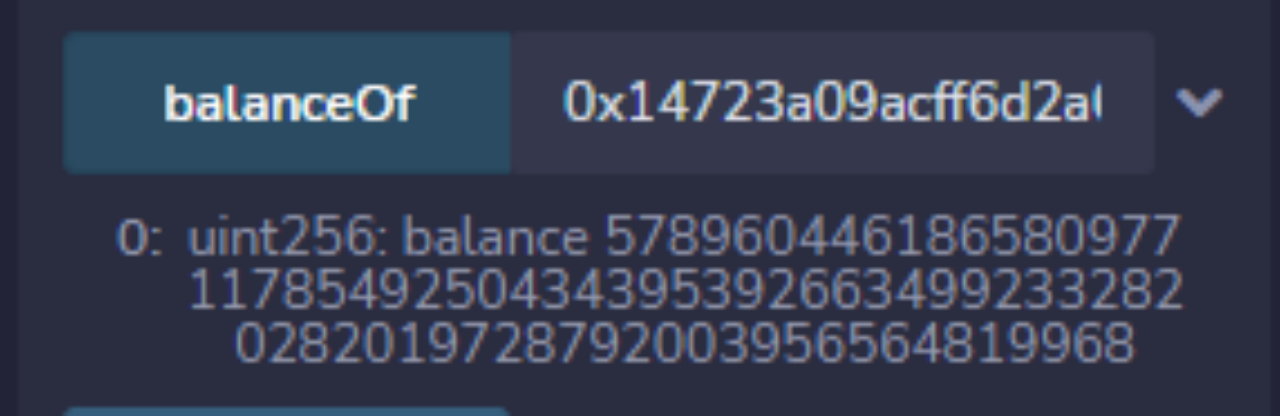}
        \caption{Abnormal Transaction Result (2)}
        \label{fig:4-21}
    \end{minipage}
    \vspace{-1em}
\end{figure}

During exceptional operations, the construction of the value parameter and the number of transaction addresses causes an overflow to 0, bypassing the condition check. As a result, during the transfer, the balance is reduced by the total transfer amount, which overflows to 0, leaving the balance unchanged. However, during the distribution, the value is set to its original value, causing a significant increase in the recipient account's balance. Figure \ref{fig:4-20} and Figure \ref{fig:4-21} show the result of the abnormal transfer transaction.

  Under normal operations, the contract deducts the correct amount from the sender and adds it to the recipient. However, by crafting specific values for \texttt{value} and recipient addresses, an overflow is triggered, setting the internal calculated result to zero. This bypasses the balance check. When performing the transfer, the subtraction causes an underflow, leaving the sender's balance unchanged. Meanwhile, the value is transferred in full, causing a significant increase in the recipient's balance.

\paragraph{\textbf{(2) Secure Mode}}

  \subparagraph{(i). Contract deployment and initial setup.}

\begin{figure}[H]
      \centering
      \includegraphics[width=0.9\linewidth]{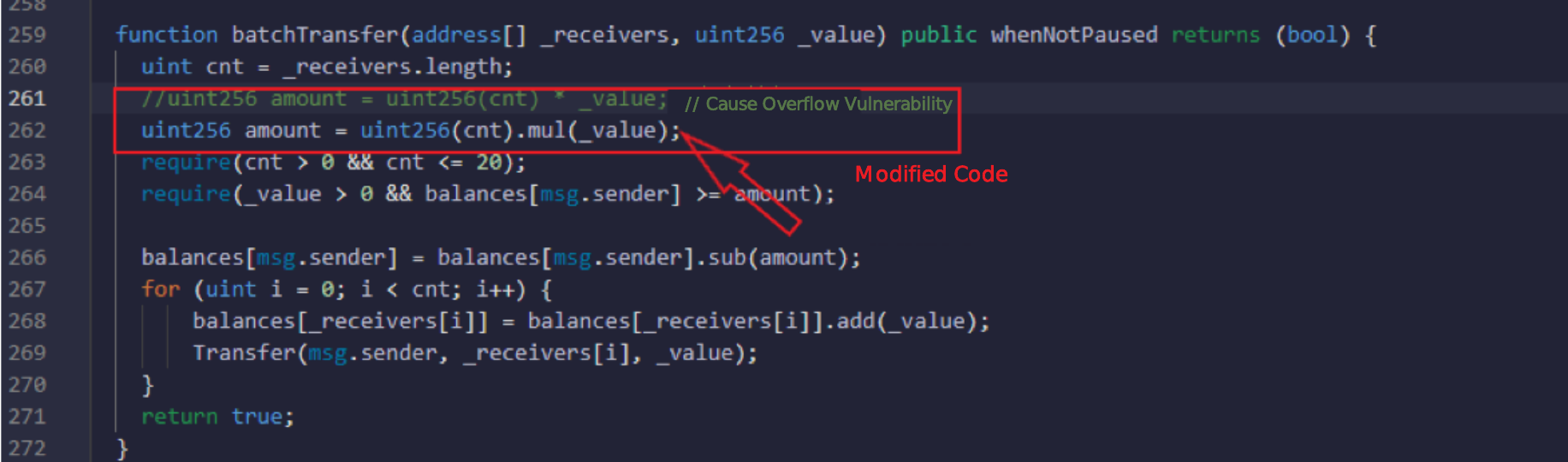}
      \vspace{-1em}
      \caption{Secure Contract Deployment Using SafeMath}
      \label{fig:4.22}
      \vspace{-1em}
  \end{figure}

Figure \ref{fig:4.22} shows the deployment of the contract in secure mode.

  \subparagraph{(ii).Attack Results.}

\begin{figure}[H]
      \centering
      \includegraphics[width=0.9\linewidth]{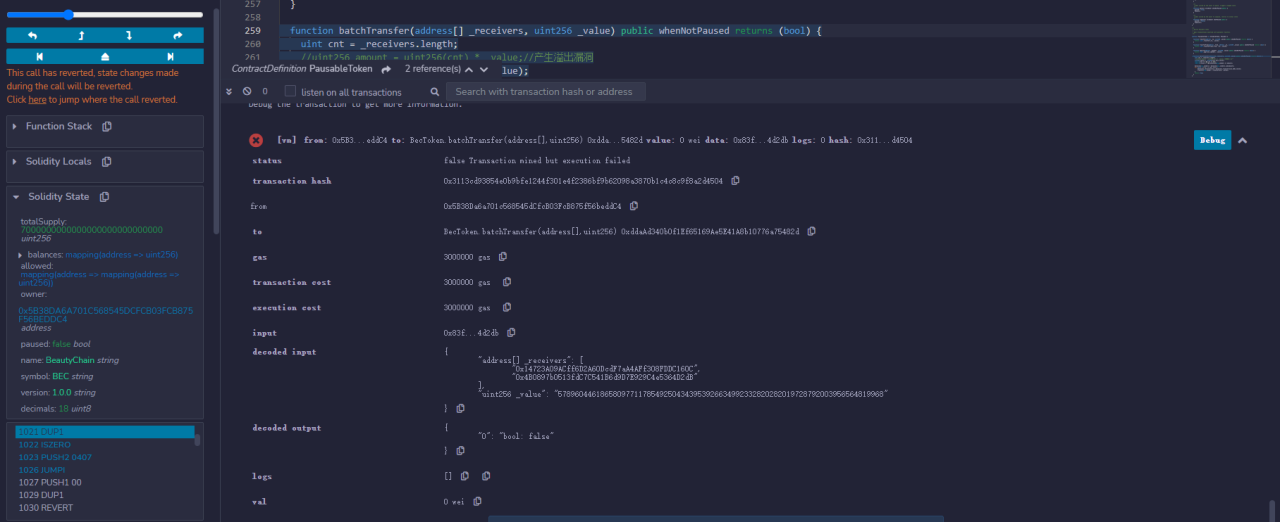}
      \vspace{-1em}
      \caption{Attack Outcome Using SafeMath}
      \label{fig:4.23}
      \vspace{-1em}
  \end{figure}

  As shown in Figure \ref{fig:4.22} and Figure \ref{fig:4.23}, modifying the code to use SafeMath, the overflow operation fails, an exception is thrown, and execution is halted.

  \subparagraph{(iii). Result analysis.}

  SafeMath uses \texttt{assert()} statements to check for arithmetic anomalies. Any attempt to exceed the bounds of integer operations will revert the transaction, effectively preventing overflow vulnerabilities.

\vspace{1em}

\section{Conclusion}



To mitigate reentrancy vulnerabilities caused by unbounded gas forwarding in \texttt{call()} and the malicious exploitation of fallback functions, employing \texttt{transfer()} for fund transfers offers a more secure alternative due to its fixed gas stipend. In the case of integer overflows, introducing arithmetic checks and adopting libraries such as SafeMath—with \texttt{assert()} to raise exceptions—effectively prevents such errors. Whether vulnerabilities stem from flaws in smart contract logic or the underlying EVM execution, ensuring smart contract security demands a holistic and multi-layered approach. This encompasses secure contract design, rigorous implementation and testing, robust regulatory oversight, and responsible user behavior. Ultimately, safeguarding smart contracts requires a collaborative effort among developers, auditors, users, and regulators to uphold both technical rigor and ethical standards throughout the contract lifecycle.



\bibliographystyle{unsrt}
\bibliography{main}

\appendix









\end{document}